\documentclass[11pt]{article}

\pdfoutput=1

\usepackage{graphics}
\usepackage{amsmath}
\usepackage{cite}
\usepackage{colordvi}
\usepackage{epsfig}
\usepackage{latexsym}

\newcommand{\slashchar}[1]{\setbox0=\hbox{$#1$}   
   \dimen0=\wd0                                     
   \setbox1=\hbox{/} \dimen1=\wd1                   
   \ifdim\dimen0>\dimen1                            
      \rlap{\hbox to \dimen0{\hfil/\hfil}}          
      #1                                            
   \else                                            
      \rlap{\hbox to \dimen1{\hfil$#1$\hfil}}       
      /                                             
   \fi}                                             %

\newcommand\gev{\text{GeV}}
\newcommand\tev{\text{TeV}}
\newcommand\pb{\text{pb}}
\newcommand\fb{\text{fb}}
\newcommand\jet{\text{jet}}
\newcommand\jets{\text{jets}}

\newcommand{\ra}{\rightarrow}

\begin{document}

\pagestyle{empty}

\vspace*{1.5cm}

\begin{center}
{\LARGE\bf \mbox{Squark Flavor Violation at the LHC}}

\vspace*{1.5cm}

Graham D. Kribs$^a$, Adam Martin$^b$  and Tuhin S. Roy$^a$ \\[1cm] 

$^a$~\textit{Department of Physics and Institute of Theoretical Science, \\
University of Oregon, Eugene, OR 97403} \\[1mm]
$^b$~\textit{Department of Physics, Sloane Laboratory, Yale University, 
New Haven, CT 06520}\\[1cm]

\end{center}

\begin{abstract}

We consider the prospects for measuring squark flavor 
violation through the signal of single top production at the LHC\@.
We study this signal in the context of $R$-symmetric supersymmetry, 
which allows for large flavor violation in the squark sector,
however the results can also be generalized to the MSSM\@. 
The single top signal arises from squark pair production 
in which one squark decays to a top and gaugino, whereas the 
other squark decays to a non-top quark and gaugino.  
We study three decay patterns:
(I)  squark decay into a quark and neutralino LSP;
(II) squark decay into a quark and neutralino NLSP, with
subsequent decay of the NLSP to a photon and gravitino;
(III) squark decay into a quark and chargino NLSP, with
subsequent decay of the NLSP to a $H^\pm/W^\pm$ and gravitino.
Case II is the most promising, when the NLSP decay is
prompt, since every event contains two hard photons
that can be used to tag the events, reducing the background
to a negligible level.
Case I is promising if the neutralino LSP is bino-like.
We carefully consider large SM backgrounds and
identify a series of cuts to isolate the signal.
Case III can occur in the minimal $R$-symmetric supersymmetric standard model (MRSSM) with Higgsino-like
lightest gauginos.  Due to the large Higgs coupling,
squarks preferentially decay to top quarks,
substantially reducing the potential flavor violating signal.
Nevertheless, the flavor violating signal might still be
identifiable if the chargino NLSP is long-lived.

\end{abstract}

\newpage

\pagestyle{plain}

\section{Introduction}

The potential misalignment of flavor between supersymmetry breaking 
parameters and the ordinary CKM matrix leads to the possibility of 
supersymmetry-induced flavor violation \cite{Ellis:1981ts,Donoghue:1983mx}.
In general the size of this flavor violation, within the 
minimal supersymmetric standard model (MSSM), far exceeds
the bounds from the myriad observations and searches for
flavor violation in the quark and lepton sector \cite{Gabbiani:1996hi}.
Constraints on flavor violation are strongest 
between the first and second generation, though substantial constraints
also exist between the third generation and the light generations.
The top sector is the least constrained, but also the most
interesting for collider physics, since tops can be 
identified given their kinematics and decay pattern.

Recently, a new approach to weak scale supersymmetry that
incorporates an (exact or approximate) $R$-symmetry \cite{Kribs:2007ac}, 
suggests large flavor violation in the supersymmetry breaking 
parameters may be present \emph{without} exceeding the 
flavor-violating bounds \cite{Kribs:2007ac,Blechman:2008gu}. 
This is possible for several reasons:
left-right squark and slepton mixing is absent;
the gaugino masses $M$ can be naturally $4\pi/g$ heavier 
than the scalar masses; and several flavor-violating
operators are more suppressed than in the MSSM due to the absence 
of $R$-violating operators.
In this paper, we study one particularly interesting collider 
signal of flavor violation within the context of an
$R$-symmetric model:  single top production from squark 
pair production in which one squark decays to a top and
gaugino, while the other squark decays to a non-top quark and
gaugino.  Interestingly, another source of single top can result 
from sgluons in $R$-symmetric models, as was recently shown 
by \cite{Plehn:2008ae}.

Single top is an important process in the Standard Model (SM) 
\cite{Bernreuther:2008ju}.  
Beyond the Standard Model, in the context of supersymmetry,
several groups have explored a single top signal with or without
additional flavor violation in the squark mass matrices
\cite{%
Tait:2000sh,%
Liu:2004bb,%
Guasch:2006hf,%
Eilam:2006rb,%
Cao:2007dk,%
Bozzi:2007me,%
LopezVal:2007rc,%
Bejar:2008ub}
as well as resulting from $R$-parity violation
\cite{Datta:1997us,Oakes:1997zg,Berger:1999zt,Berger:2000zk}.
Unlike the approaches in these papers, however, our single top signal
does not suffer from substantial restrictions from other 
flavor-violating processes.  The presence of an $R$-symmetry
suggests large flavor-violation can be probed in a variety of ways 
involving sleptons as well as squarks.  Nevertheless, we make no 
attempt at an exhaustive study of flavor violation.

Our focus is on the shortest possible decay chains of squarks: 
pair-produced squarks that decay into different flavors of quarks
and the lightest gaugino.  We consider three basic scenarios:
(I)   the ``collider-equivalent'' lightest supersymmetric particle
      (LSP) is a neutralino,
(II)  the next-to-LSP (NLSP) is a neutralino that decays within 
      the detector to a photon and a gravitino LSP, and
(III) the NLSP is a chargino that may or may not decay (within the detector)
      to a gravitino LSP.

Cases II and III arise when the scale of mediation of 
supersymmetry breaking is low \cite{Giudice:1998bp} 
or as a consequence of a strongly coupled 
hidden sector \cite{Cohen:2006qc,Roy:2007nz,Murayama:2007ge}.   
In an $R$-symmetric model, $R$-symmetry itself may be broken 
explicitly in the hidden sector to cancel the cosmological constant.  
This explicit breaking, communicated via anomaly-mediation to the 
visible sector, 
leads to suppressed contributions to $R$-violating supersymmetry 
breaking parameters.  The size of this contribution is proportional
to the gravitino mass, which cannot exceed the weak scale to
ensure the $R$-violation in the visible sector, 
proportional to $\alpha/4 \pi \, \times$ 
the gravitino mass, is sufficiently small.
Given a much smaller mass for the gravitino, where it 
becomes the LSP, the induced $R$-violation automatically becomes safe.
Two interesting scenarios for a signal of flavor violation result
within the MRSSM with a gravitino LSP:
one with a neutralino next-to-lightest supersymmetric
particle (NLSP); the other, a chargino NLSP\@.  
Depending on the strength of the interaction between the NLSP and the LSP,
the NLSP can be long lived. The characteristic signals depend
sensitively on the lifetime of the NLSP\@.

Finally, our MRSSM analysis can also be applied to the 
MSSM in a partially $R$-symmetric limit, which may well be 
of interest in its own right.  As a reminder, $R$-symmetry forbids 
left-right mass mixings among squarks and sleptons, 
and also implies gauginos are Dirac fermions.
(A brief review of the MRSSM and its characteristics is 
provided in the Appendix of this paper.)
At the LHC, squark two-body decay into a Dirac gaugino LSP 
is indistinguishable from squark decay two-body decay into 
a Majorana gaugino LSP\@.  Then in the MSSM, left-right mixing 
is proportional to $m_f (A_f - \mu/\tan\beta)$ or $m_f (A_f - \mu \tan\beta)$ 
for up-type or down-type sfermions, and thus can vanish 
for particular choices of parameters.  It is in this limit
that our results apply to the MSSM\@.

The organization of this paper is as follows.  
In Sec.~\ref{sec:threecases} we outline the three classes of signals
that we consider in this paper.  We then proceed to analyze
each case in subsequent sections.  
In Sec.~\ref{sec:case1}
we consider the signal top signal with a neutralino LSP\@.
In Sec.~\ref{sec:case2}
we consider the signal top signal with a neutralino NLSP
that decays within the detector to a gravitino and photon.
In Sec.~\ref{sec:case3}
we consider the signal top signal with a chargino NLSP\@.
Finally, we conclude in Sec.~\ref{sec:conclusions}.


\section{Squark Flavor Violation:  3 Cases of 3 Signals}
\label{sec:threecases}

We are concerned exclusively
with squark production and decay at colliders, specifically
at the LHC\@.  Consequently, for us ``LSP'' always refers to 
``collider-equivalent LSP''.  That is, the last particle in
the supersymmetry decay chain that escapes the detector.
This means that we treat a strict neutralino LSP the same as
a neutralino collider-equivalent LSP where the latter 
decays into a gravitino well outside the detector.

There are three separate scenarios with a signal of 
squark flavor violation that we focus on in this paper:
\begin{itemize}

\item[I:]  The lightest neutralino is a collider-equivalent LSP\@. The
  shortest decay chain possible is for the squark to decay directly
  into a quark plus neutralino.  Within this class of processes, we examine 
  the flavor-violating signal of a single top, {\it i.e.}, we
  select events that contain one top quark and one quark of different
  flavor. Top is identified
  by the detection of its decay products ($b$ and a leptonic decay of $W$). 
  The specific collider signal is thus one lepton, one tagged $b$ jet,
  and one other jet. 
  We look for flavor violation in single top for three
  reasons: first, flavor violation in the up sector, especially flavor
  violation involving the third generation, is relatively
  unconstrained. Second, the hard lepton from $W$-decay provides a
  good trigger for these events and an effective way to reduce the
  immense QCD background. Lastly, $b$-jets can be tagged most efficiently among all jets.

\item[II:] The lightest neutralino is the NLSP, which decays 
  within the detector to a gravitino LSP\@.\footnote{Even if the 
  neutralino decay is not prompt, as long as the decay
  happens before the electromagnetic calorimeter, the resulting photon
  may be observed.}
  The NLSP decays into a gravitino and a photon or a $Z$.
  Decays to a $Z$ are typically kinematically suppressed compared to
  the decays to 
  a photon, and thus we will focus on signals containing two hard
  photons plus missing energy.  For supersymmetric single-top
  production, the parton-level final state is now
  $bj\ell\gamma\gamma + \slashchar{E}_T$. 
  However, because the hard photons provide a reliable
  trigger and excellent background discrimination, we do not
  require a lepton to suppress SM background in this scenario. 
  Therefore we are free to
  consider flavor-violating processes other than single-top
  production: For example, we may also look at flavor violation in
  the sbottom sector $pp \rightarrow \tilde q_d \tilde q^*_d
  \rightarrow b + d + \chi_1 \overline{\chi}_1$, in which case
  the final state of interest is $b j \gamma\gamma + \slashchar{E}_T$.

\item[III:] The chargino is the NLSP, and the gravitino is the LSP\@.
  The main difference from Case II is that if the chargino 
  is long-lived, it produces tracks inside the detector. 
  Accordingly, three different sub-cases 
  arise depending on the decay length of the chargino: 
  \begin{itemize}
  \item[III.a:] The decay of the chargino is prompt.  In this case,
    the single top signal arises from down-squark pair production
    into one $t$ (decaying into a $b$ and $W$) and one jet with
    two $H^\pm/W^\pm$.  Up-squark pair-production gives a similar
    signal, except that the $b$ is produced directly, and thus
    there is no additional $W$ from top decay.
  \item[III.b:] The chargino is long-lived, but decays inside the
    detector. 
  \item[III.c:] The chargino decays outside the detector, producing 
    tracks of charged heavy particles escaping the detector.   
   \end{itemize}

\end{itemize}

In the next few sections we systematically analyze these cases
in the context of observing the single top signal of squark 
flavor violation.

\section{Case I:  A Neutralino LSP}
\label{sec:case1}

\subsection{Flavor Violation in Single Top: Setup and Feasibility}
\label{subsec:decayratio}
Single top production is widely understood to be an
important process as it allows a direct test of the unitarity of 
the CKM matrix.
In the SM, however, single top production proceeds only through 
electroweak interactions and, depending on the production channel, 
it is also suppressed by the $b$-quark PDF, CKM angles, or
phase space.
Thus, this provides an opportunity for a beyond the SM signal 
to be seen in this channel at the LHC\@. 
In our scenario, the BSM-induced single top signal can receive 
significant enhancement due to the flavor-violating elements in the 
scalar mass matrices in the MRSSM\@.  Our goal is to isolate this
signal from among the SM and detector backgrounds.

The signal events arise due to the following processes in the MRSSM 
\begin{equation}
  \label{eq:process1}
  p + p \ \rightarrow  \ \left\{ \ 
   \begin{matrix}  \tilde u_{L_a} + \tilde u_{L_a}^* \\ 
                   \tilde u_{R_a} + \tilde u_{R_a}^* 
   \end{matrix} \right\} \
      \rightarrow  \  \text{top + jet} +  \chi_1 
           + \overline  \chi_1  \
      \rightarrow  \ W + b + \text{ jet} 
           +  \chi_1 + \overline  \chi_1   \; ,
\end{equation}
where we used notation from the Appendix:  the up-type squarks
$\tilde u_{L_a}$ and $\tilde u_{R_a}$ are mass eigenstates and 
$\chi_1$ is the lightest neutralino of the MRSSM\@.  For this Case, 
$\chi_1$ escapes the detector, giving rise to additional missing
energy. We require that the $W$ decays leptonically for triggering
purposes and to suppress multijet QCD backgrounds. 

This flavor-violating signal requires that the two squarks 
decay differently -- one decays to top, and the other to 
an up or charm quark.  Both decay modes must have significant 
branching ratios. 
A necessary condition for a significant number of signal events is then
the presence of large mixing angles in the squark mass matrices. In
order to simplify the discussion, we take the squark mixing matrix 
to be
\begin{equation}
  \label{eq:mix1}
  U_{\tilde u_L} = \begin{pmatrix}
                  \cos \theta_L & 0 & \sin \theta_L \\
                    0 & 1 & 0 \\
                  - \sin \theta_L & 0 & \cos \theta_L  
                 \end{pmatrix} , \qquad
  U_{\tilde u_R} = \begin{pmatrix}
                  \cos \theta_R & 0 & \sin \theta_R \\
                    0 & 1 & 0 \\
                  - \sin \theta_R & 0 & \cos \theta_R  
                 \end{pmatrix} \qquad
\end{equation}
so that we have mixing between two species of the squarks only. 
Sfermion mixing matrices are defined in App.~\ref{app:matter},
while the Lagrangian written in the mass eigenstate basis 
incorporating these mixing matrices can be found
in App.~\ref{app:lagrangian}.
In the following, we will assume the mixing angle to be maximal.
The size of our signal, however, can be simply scaled with 
the mixing angle given otherwise identical kinematics.

The presence of a large mixing angle, by itself, is not sufficient
to yield the flavor-violating signal we seek.  The squark masses 
also play a crucial role - they determine whether or not decay 
channels are kinematically open.  Throughout this paper we 
assume squark masses are large enough such that decays to top and 
the lightest neutralino is not kinematically forbidden.  More importantly, 
the branching ratios depend crucially on the composition of 
$\chi_1$, i.e., how much bino, neutral wino, and neutral Higgsino 
is present.  There are thus three distinct limits: 

\begin{itemize}

\item[a:]  The lightest neutralino is mostly a bino. Such a
  scenario arises when $M_1 \ll \mu_u, \mu_d, M_2$.\footnote{As an aside, 
  this is precisely the mass ordering that was needed in 
  Ref.~\cite{Harnik:2008uu} for a Dirac bino to explain the 
  PAMELA positron ratio excess \cite{Adriani:2008zr}.}
  Here $M_1, M_2, \mu_u, \mu_d$ are the $R$-symmetric bino, wino, 
  up-type Higgsino and down-type Higgsino masses, defined in 
  App.~\ref{app:gaugehiggs}.
\item[b:] The lightest neutralino is mostly a neutral Higgsino.
  This occurs if either or both of {$\mu_u,\mu_d$} are much smaller 
  than $M_1$ and $M_2$. 
\item[c:] The lightest neutralino is mostly a wino.  This occurs
  when $M_2 \ll \mu_u, \mu_d, M_1$.  This scenario is disfavored 
  when $M_2 < 1$~TeV, since a vev develops for the scalar SU(2) triplet, 
  which leads to 
  an excessive contribution to $\Delta \rho$ \cite{Kribs:2007ac}.
  Also, a light wino in the presence of flavor violation in the up sector 
  also could lead to sizable contributions to FCNCs such as 
  $K^0-\overline{K}^0$ mixing.  We do not consider this possibility
  further.

\end{itemize}

The difference in the first two limits can be easily seen in the
gaugino-squark-quark couplings given in the Appendix ({\it c.f.},
Eq.~\eqref{nut-eq-ew}; in this equation, neutralinos are in 
electroweak gauge eigenstates). The only quark which couples to up-type squarks in the squark-quark-Higgsino interactions is the top quark. This
implies that, if the   
lightest neutralinos are Higgsinos (in the notation of the Appendix, $\bar N P_3$ and $\bar N P_4$), the two body decay of up-type squarks to the lightest neutralino always proceed to a top quark -- regardless of the squark flavor content. This happens
precisely since we consider only the third generation quark masses
to be nonzero.  Even if the masses of the first two generations are
taken into account, the decay of a squark into a Higgsino and a lighter 
quark is highly suppressed when compared to the decay into top.  
Hence, finding flavor-violation with a Higgsino-like lightest
gaugino is extremely challenging.

\begin{figure}[t]
\centering
\begin{minipage}[c]{0.45\linewidth}
   \centering
   \includegraphics[width = 2.9 in]{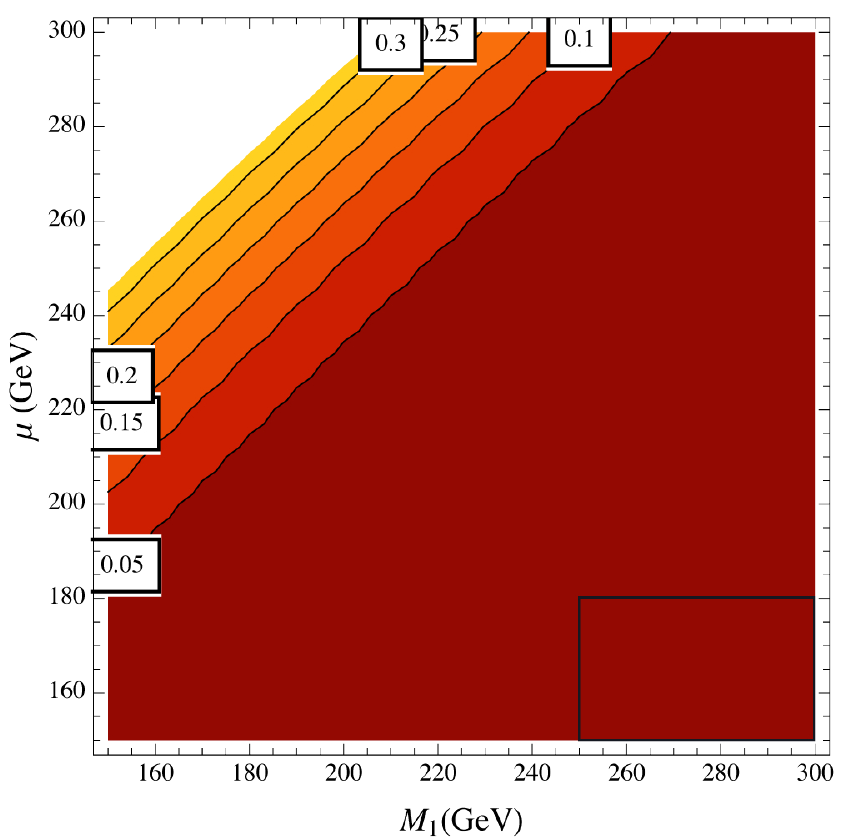} 
\end{minipage}
\hspace{0.5 cm}
\begin{minipage}[c]{0.45\linewidth}
   \centering
   \includegraphics[width = 2.9 in]{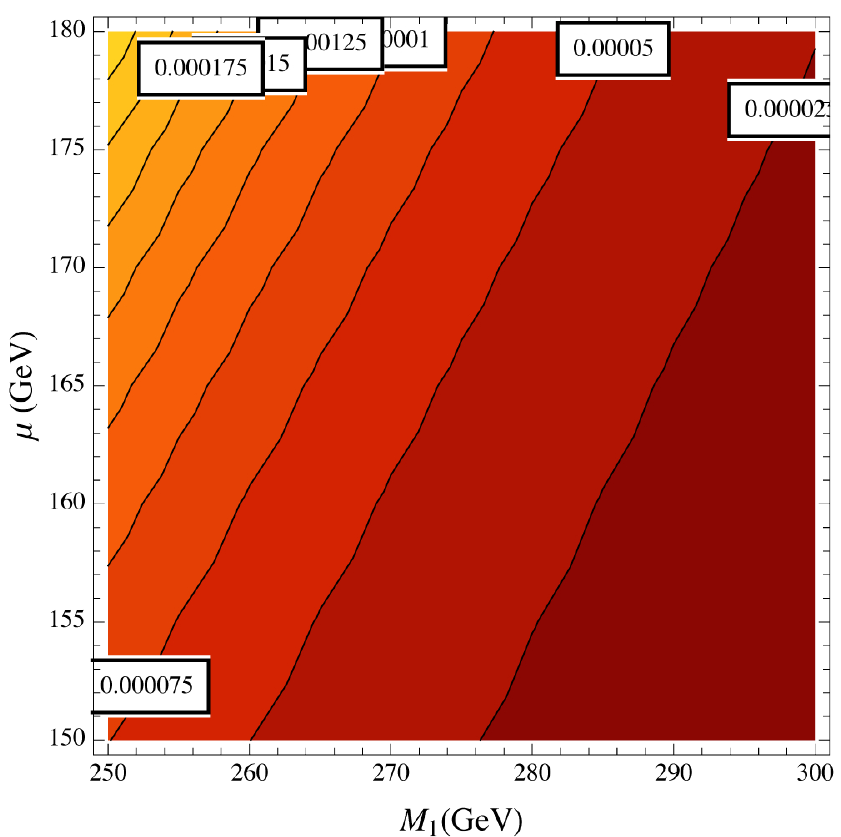}
\end{minipage}  
\caption{Contours of $r_{\tilde u_L}$ in the MRSSM for $\tan\beta=10$
  and $M_2 = 1$~TeV\@. The rightmost figure is a magnification of the rectangular region indicated in the bottom-right corner of the left figure. It shows the rapid decrease of $r$
  with increasing $M_1/\mu$.} 
\label{fig:1}
\end{figure}

\begin{figure}[t]
\centering
\begin{minipage}[c]{0.45\linewidth}
   \centering
   \includegraphics[width = 2.9 in]{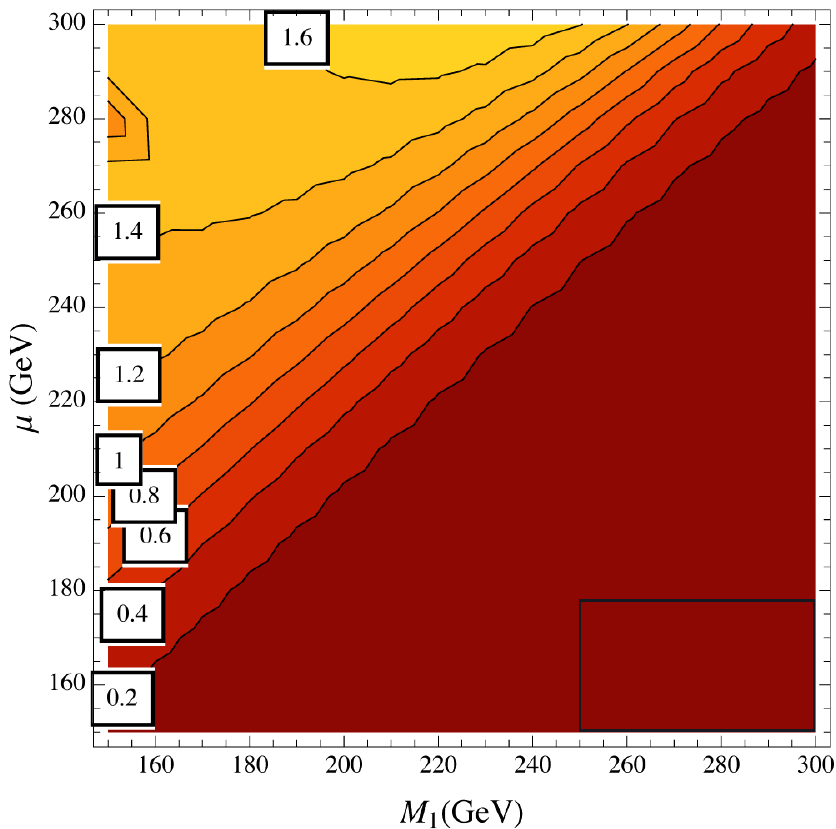} 
\end{minipage}
\hspace{0.5 cm}
\begin{minipage}[c]{0.45\linewidth}
   \centering
   \includegraphics[width = 2.9 in]{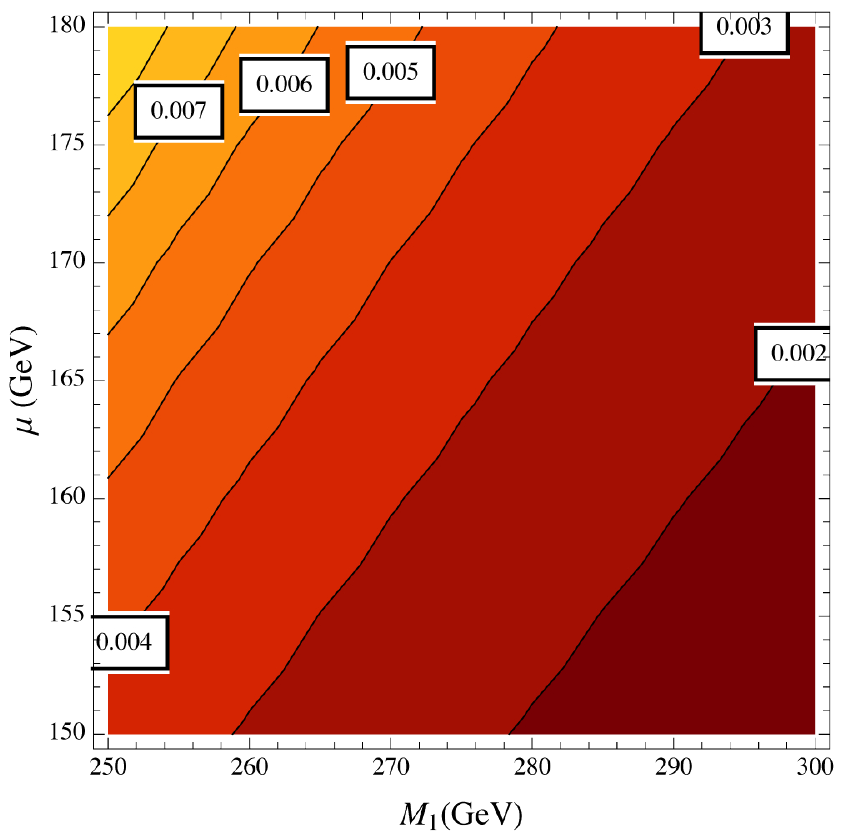}
\end{minipage}  
\caption{Contours of $r_{\tilde u_R}$ in the MRSSM for $\tan\beta=10$
  and $M_2 = 1$~TeV\@.  The rightmost figure is a magnification of the rectangular region indicated in the bottom-right corner of the left figure. It shows the rapid decrease of $r$
  with increasing $M_1/\mu$.}
\label{fig:2}
\end{figure}

The bino limit is much more promising.  Gauge invariance
requires that the couplings are universal, independent of generation
(in the pure bino limit).  The only difference arises due to
squark mixing angles.  It is useful to define the flavor-violating
ratio $r$ 
\begin{equation}
  \label{eq:R-def}
  r_a \equiv \frac{\Gamma_{\tilde q_a \rightarrow u \chi_1}}
      {\Gamma_{\tilde q_a \rightarrow t \chi_1}} 
\end{equation}
which characterizes the amount of up-type flavor violation
revealed in squark decay.  To illustrate how large $r$ can be
within the MRSSM parameter space, in Figs.~\ref{fig:1},\ref{fig:2}
we have plotted $r$ as a function of $M_1$ and $\mu_u = \mu_d = \mu$
where we took $\theta_L = \theta_R = \pi/4$ and $M_2 = 1$~TeV\@. 
Maximal mixing angle implies that each squark mass eigenstate 
couples equally to $t$ and $u$.  Taking the mass of the squark to be 
$500$~GeV, in Figs.~\ref{fig:1},\ref{fig:2} we plot $r$ for $\tan
\beta = 10$. 


Clearly, the region with $r \sim 1$ occurs when the 
lightest neutralino is mostly a bino.
In addition, note that using the mixing angles shown in Eq.~\eqref{eq:mix1},
we find that the amplitude for the decay 
$\tilde u_{L_1}+ \tilde u_{L_1}^* \rightarrow t + \bar{q}_i + \chi_1
\bar \chi_1$ is proportional to $\sin 2 \theta_L$, whereas the decay
of $\tilde u_{L_3}+ \tilde u_{L_3}^*$ to the same final states is
proportional to $ - \sin 2 \theta_L$. This implies that, if these two
mass eigenstates are of equal mass, then there is complete cancellation
of the amplitudes (a squark GIM mechanism). Thus, in order to have a non-zero MRSSM
contribution to single
top, the scalar masses must be hierarchical. 
Finally, the gaugino hierarchy $M_1 \ll \mu_u, \mu_d, M_2$
automatically ensures that all charginos and other neutralinos 
are also heavy.  If the squarks are light enough, it is plausible
that the only open channel for the two body decay of a squark is
into a $\chi_1$ and quark.


\subsection{Signal and Background: A Quantitative Analysis}

There are considerable standard model events which constitute the
background for single top events in Eq.~\eqref{eq:process1} due to
flavor violation. One needs to 
eliminate not only all the electroweak single top events but also all the standard model events which
mimic single top.
A quantitative study that explores the feasibility of finding 
flavor violation at LHC in the new physics production of single top 
must include all these backgrounds. In this subsection we systematically
analyze signals due to sample spectra on top of the standard model
backgrounds after imposing a set of cuts and show how the significance
of signal vary  as we change parameters in the spectrum.

We start this subsection with a description of the software
tools we have used. We list all the channels that have relevant
contributions towards the background along with their cross-sections
after using generic parton level cuts. We then compare  these
background events to the signal events generated in a sample spectrum
after we impose various well motivated cuts and estimate the
significance of the signal. We also show how the significance
varies as the spectrum itself is varied while keeping the same set
of cuts used before.  We end this subsection with a brief discussion of 
further refinements.

\subsubsection{Software}
\label{sec:software}

To simulate the signals and background necessary for our analysis of
flavor violation, we used several different Monte Carlo event
generators. All events were first created at parton level utilizing
matrix-element generators: ALPGENv13~\cite{Mangano:2002ea} for the
backgrounds, and 
MadGraphv4~\cite{Maltoni:2002qb,Alwall:2007st} for the signal. The
inputs to the matrix element generators 
are a set of parton level cuts, which we list under
Table~\ref{table:sigbr}, a 
factorization/renormalization scale, and a parton distribution function
(PDF) set. For
factorization/renormalization scales, we used the default ALPGEN
values when generating each background, and the squark mass
($\sim 300 - 500\ \gev$) for the signal. The default PDFs were used
throughout -- CTEQ5L for the background and CTEQ6L1 for the signal. 
All parton level events (signal +
background) were passed through PYTHIA6.4~\cite{Sjostrand:2006za} for showering and hadronization, and PGS4.0~\cite{PGS} for detector simulation. We use the
parameters in the default {\tt pythia$\_$card} and {\tt
  pgs$\_$card$\_$ATLAS} provided with MadGraph for all
events.\footnote{The only modification of the PGS card is a change in
  the sagitta resolution from $10^{-5}$ to $10^{-4}$. This primarily
  effects muon resolution~\cite{Brooijmans:2008se}.} The 
primary effects of the detector simulator are: 

\begin{itemize}
\item Limited calorimeter size and threshold: Electrons and photons are
  visible for $E_T > 5$~GeV, $|\eta| < 3$. 
  Muons are visible for $E_T > 3$~GeV, $|\eta| < 2.5$.
  Jets are visible for $E_T > 5$~GeV, $|\eta| < 4$. 
\item Lepton energy smearing: ID efficiency for leptons in our
  kinematic region of interest is $80-90\%$. 
\item Jet smearing: The energy/momentum for jets are smeared by an amount
\begin{equation}
\frac{\delta E_{\jet}}{E_{\jet}} = \frac{0.8}{\sqrt{E_{\jet}/\gev}} \; .
\end{equation}
Not only does this affect the resolution of any object reconstructed
from jets, but it also affects the missing energy: $\slashchar{E}_T =
-\sum_{{\rm visible}}p_T$, so an error in the visible energy becomes
an error in the missing energy. 
\item $b$-tagging: The PGS $b$-tag efficiency peaks around $65\%$ with
  a corresponding fake rate of $\sim 1$-$2\%$. Both the tag and mistag
  rates vary with $p_T, \eta$ and are modeled in PGS using a CDF
  fit. A $b$-jet can only be tagged if it goes through the tracker
  which has pseudo-rapidity extent $|\eta| < 2.5$. 
\end{itemize}
	
\subsubsection{Backgrounds}
		
In order to avoid potentially disastrous QCD backgrounds, we
require the $W$ from top quark decay to itself decay leptonically. 
Therefore the signal we wish to identify, in terms of objects seen in the
detector, is: 
	\begin{equation}
	\label{eq:LHCstate}
	pp \rightarrow b + j + \ell + \slashchar{E}_T,\ \ell = e, \mu
	\end{equation}
An important distinction between the signal and background
is the source of missing energy.  The source of missing energy 
in the background comes from neutrinos from $W, Z$ and hadron decays, 
while the missing energy for the signal also contains the neutralinos. 
SM backgrounds for the final state given in Eq.~(\ref{eq:LHCstate}) are:
\begin{itemize} 
\item single top: $t + q \rightarrow b\ell \nu j$, where $q$ can also
  be $b$ quark. 
\item top + $W$: $t + W + \jets \rightarrow bWW + \jets$,
  where at least one $W$ decays leptonically.  
\item top pair production: both semileptonic $t\bar t
  \rightarrow b\ell \nu \bar b jj$ and fully leptonic $t\bar t
  \rightarrow b\bar b \ell \ell' \nu \nu'$ decay modes. 
\item $W\ +$ heavy flavor: $W + \bar b b \rightarrow b\bar b \ell \nu$.
\item $Z(\bar{\nu}\nu)$ + heavy flavor:  $Z+\bar b b \rightarrow b\bar
  b + \bar{\nu}\nu$ 
\item $W(\ell \nu)\ + \jets$,
\item $Z(\bar{\nu}\nu, \ell^+ \ell^-) + \jets$
\item $W(\ell \nu) + Z(\bar{\nu}\nu) + \jets$
\end{itemize}
	
A few comments on the backgrounds are in order. First, many SM
backgrounds contain only one neutrino and therefore only one source of
missing energy.  Second, several backgrounds ($t \bar t$ and $W + b
\bar b$ in particular) require one of the $b$ quarks to be
mistagged, while others ($W+\text{jets}$) require a light jet to fake
a $b$-jet. Both of these facts will be exploited in the cuts section 
to separate the signal from the background.  Information on the signal
and backgrounds, including which generator was used and what cross
section was assumed, are contained below in Table~\ref{table:sigbr}. 
 
\begin{centering}
\begin{table}[h!]
\centering 
\begin{tabular}{ |c|c|c| }\hline
Process & $\sigma $ & \# events  \\ 
\hline \hline  & & \\ 
$\tilde u_i \tilde u^*_i \: \rightarrow \: 
t j \, \chi_1 \overline{\chi}_1 \: \rightarrow \: 
\ell b j \nu \, \chi_1 \overline{\chi}_1$  
& $1.88\ \pb$ & $14,300$  \\ & &  \\
\hline  \hline  & & \\
$t\bar t \: \rightarrow \: b\bar b j \, \ell \nu $ & $197\ \pb$ &  
$2.90\times 10^6$ \\
$t\bar t \: \rightarrow \: b\bar b\ell \ell' \nu \nu'$ & $49.1\ \pb$ & 
$2.09\times 10^6$ \\
$t + q \: \rightarrow \: bj \ell \nu$ & $59.2\ \pb$ & 
$1.8\times 10^6$ \\ 
$t + b \: \rightarrow \: b b \, \ell \nu$ & $2.28\ \pb$ & 
$1.12\times 10^6$ \\  
$t(\text{inc.}) + W(\ell\nu)$ & $17.9\ \pb$ & 
$1.01\times 10^6 $ \\  
$t(\text{inc.}) + W(\ell\nu) + j $ & $31.3\ \pb$ & 
$9.41\times 10^5$ \\   
$W + \bar b b \: \rightarrow \: \bar b b \ell \nu$ & $17.6\ \pb$ &
$8.75\times 10^5$ \\ 
$Z + \bar b b \: \rightarrow \: \bar b b \bar{\nu} \nu$ & $24.7\ \pb$ &
$6.30\times 10^5$ \\ 
$W Z j j \: \rightarrow \: \ell + 3 \nu + j j$ & $1.23\ \pb$ & 
$2.1\times 10^5$ \\  
$W + \text{ jets} \: \rightarrow \: \ell \nu + \text{ jets} $ &  &  \\
$W + j$ & $525.3\ \pb$ & $1.21\times 10^6$ \\
$W + jj$ & $744.5\ \pb$ & $6.40\times 10^6$ \\
$W + \text{ more than } 2j$ & $396\ \pb$ & $3.58\times 10^6$ \\
$Z + \text{jets}\: \rightarrow \: \ell^+\ell^- + \text{ jets}$ & & \\
$Z + j $ & $1737\ \pb$ & $1.27\times 10^7$ \\
$Z + jj $ & $967\ \pb$ & $6.4\times 10^6$ \\
$Z + \text{ more than } 2j$ & $291\ \pb$ & $2.9\times 10^6$ \\ 
$Z + \text{jets}\: \rightarrow \: \bar{\nu}~\nu+ \text{ jets}$ & & \\
$Z + j $ & $79\ \pb$ & $3.49\times 10^6$ \\
$Z + jj $ & $439\ \pb$ & $4.24\times 10^6$ \\
$Z + \text{ more than } 2j$ & $229\ \pb$ & $2.89\times 10^6$ \\ 
\hline
\end{tabular}
\caption{Signals and background processes considered. Cross
  sections include the branching ratios to leptons ($\ell = \mu,
  e$). All events were generated with parton level cuts of $p_{T,j},
  p_{T,l} > 15.0~\gev,\ |\eta_j|, |\eta_b| < 4.0$, and $\Delta R_{jj}
  > 0.4$, while the $W+ \text{jets}$ and $Z(\bar{\nu}\nu) + \text{jets}$ backgrounds were generated with the
  additional cut of $\slashchar{E}_T> 75.0~\gev$. Jet-parton matching, following the MLM scheme~\cite{Catani:2001cc}
 was used in all backgrounds except for single-top, where it is not yet available, and $W/Z + $ more than $2$ jets. The parton level
  cuts were imposed to make event generation 
  more efficient; ideally parton level cuts should be softer then the
  analysis cuts to allow initial and final state radiation to have an
  effect. We used the default factorization and renormalization
  schemes/scales for all of the above events. The number of events for
  the background was chosen to 
  roughly correspond to the number of events expected (after the full
  analysis) after $10\ \fb^{-1}$ of integrated luminosity.} 
\label{table:sigbr}
\end{table}
\end{centering}

\subsubsection{A Sample Spectrum}
\label{sec:sample}

In order to produce signal events, we use the following MRSSM
spectrum: 
\begin{gather*}
 m_{\tilde u_{L_1}}  =  m_{\tilde u_{R_1}} = 1~\tev,\qquad
 m_{\tilde u_{L_3}} = 1~\tev,\quad  m_{\tilde u_{R_3}} = 300~\gev, \\
 M_1 = 50~\gev, \qquad M_2 = 1~\tev, \qquad M_3 = 3~\tev, \\
\mu_u = \mu_d = 1~\tev, \qquad \text{ and} 
\quad \tan{\beta} = 10 \\
\theta_R  = \pi/3,\qquad  \theta_L = 0\; . 
\end{gather*}
This is the simplest set of MRSSM parameters which has all of the
attributes suggested 
previously for a large flavor-violating signal: the lightest
neutralino is primarily a bino, there is a hierarchy between the
masses 
of the squarks which mix among themselves, and all squarks are
heavy enough so that the decay channel to the LSP + top is open. 

The gluino and all squarks except one right-handed up-type squark, 
$\tilde u_{R_3}$ are heavy, thus $\tilde u_{R_3}$ completely 
dominates SUSY production at the LHC\@. We chose $\tilde u_{R_3}$ 
to be light (rather than $\tilde u_{L_3})$ as it is an electroweak singlet, so its mass can be 
adjusted without affecting the other states in the spectrum. 
Since $\mu_d, \mu_d, M_2  > m_{\tilde u_{R_3}}$, all the charginos 
and all the neutralinos except the LSP are heavier than the 
$\tilde u_{R_3}$ squark.  The only open two-body decay channels 
are $\tilde u_{R_3} \rightarrow \chi_1 + t$, 
$\tilde u_{R_3} \rightarrow \chi_1 + u$ both of which involve the 
weak bino coupling, hence $\tilde u_{R_3}$ is very narrow, 
$\Gamma_{\tilde{\chi}} \sim \mathcal O(1\ \gev)$.  
By restricting the $\tilde u_{R_3}$ decays to $q + \tilde{\chi}_1$ 
we enhance the cross section for the flavor-violating 
signal, however this comes at a price: we have fewer tools to fight
the SM background, and our signal cannot provide much information on
the details of the MRSSM spectrum. To get detailed spectrum
information a scenario which has  smaller cross section but contains
long SUSY decay chains, similar to those considered in Ref.~\cite{Athanasiou:2006ef, Wang:2006hk, Cheng:2007xv, Cheng:2008mg, Burns:2008va},
would likely be better. 

\subsubsection{Cuts}
\label{singletop-cuts}

To suppress the SM background we impose the following cuts:
\begin{itemize}
\item Exactly $1$ lepton, $p_T> 30~\gev, |\eta| < 2.5$. This serves
  as our primary trigger and suppresses QCD backgrounds. 
\item Exactly $2$  $\jets$, within $|\eta| < 3.5$. One jet
must be tagged as a $b$-jet and have $p_T > 50\ \gev$, while the
untagged jet must have $p_T > 80\ 
  \gev$.  
\item $\slashchar{E}_T > 100~\gev$
\item $\slashchar{E}_T > 0.25\times M_{eff}$, where $M_{eff} =
  \sum_{j, \ell}p_T + \slashchar{E}_T$ 
\item Transverse mass of the $W$, $m_{T,W} > 120\ \gev$
\end{itemize}
These cuts were not optimized in any rigorous way.

The motivation for these cuts is the following: First, for a process
where the sole source of (true) missing energy is the 
neutrino from $W$ decay, the transverse mass of the
lepton-$\slashchar{E}_T$ system ($m_{T,W} = \sqrt{(E_{\ell}+
  \slashchar{E}_T)^2 - (p_{T, \ell} + p_{T,\nu})^2}$) exhibits a sharp
edge at $M_W$, at least up to detector resolution effects. 
If there are multiple sources of missing
energy, as in the MRSSM flavor-violating signal, the transverse mass
distribution is much smoother. Thus, by cutting on $m_{T,W} \gg M_W$,
we strongly suppress all backgrounds except fully leptonic $t\bar t, t
W (2\ell 2\nu b), Z(\bar{\nu}\nu) + \bar b b, {\rm and}\ WZ(3\ell +
\nu) + \jets$. Of the remaining backgrounds, fully leptonic $t \bar t$
is the largest, despite the fact that it matches the signal only when
one of the leptons is missed or falls outside the calorimeter. To
reduce the residual leptonic $t\bar t$ background, we exploit the fact
that  the $\slashchar{E}_T$  
and non b-jet $p_T$ are typically higher for the signal than for 
the background.  The jet $p_T$ is higher for the signal since it
arises from a massive squark decaying to two essentially massless
objects. The heavier the squark, the harder this jet will be. The
missing energy is also larger for the signal simply because there are more
sources -- two neutralinos and a neutrino for the signal, compared to
just two neutrinos for the background. The optimum value for the
$\slashchar{E}_T$ cut also depends on the 
mass of the squark, with more massive squarks able to deposit more
missing energy. 

The other cut, exactly two jets, is designed to remove the remaining 
semileptonic $t \bar t$ and $W/Z + \text{jets}$ backgrounds. 
These processes have such large cross sections
that they pollute our signal if we only impose the $m_{T,W}$
cut. Also, by adding this cut we eliminate several MRSSM
backgrounds. The more jets we allow, the more MRSSM processes can
contribute, obscuring the interpretation of an excess as flavor
violating new physics. The 
signal and background efficiencies under these final cuts, and the
number of events generated in $10\ \fb^{-1}$ of integrated luminosity
are summarized in Table~\ref{table:eff}. 

\begin{table}[h!]
\centering 
\begin{tabular}{ | c | c | c | }   \hline 
Process & Efficiency $\epsilon$ & \# events in $10\ \fb^{-1}$  \\ 
\hline \hline & & \\
$\tilde u_i \tilde u^*_i \: \rightarrow \: t j \chi_1 \,
  \overline{\chi}_1 \: \rightarrow \: \ell b j \nu \, \chi_1
    \overline{\chi}_1$ &  0.010 & 196  \\  
 ~ $m_{{\tilde u}_R} = 300\ \gev, ~ m_{\tilde{\chi}_1} = 50\ \gev$ & & \\
  & & \\   \hline \hline
$t \bar t \: \rightarrow \: b\bar b j \ell \nu $ & 
  $5.85\times 10^{-6}$ & 12 \\ 
$t\bar t \: \rightarrow \: \ell \ell' \nu \nu'$ & 
  $5.20 \times 10^{-4}$ & 256 \\ 
$t + q \: \rightarrow \: b j \ell \nu$ &
  $7.16\times 10^{-6}$ & 4 \\ 
$t + b \: \rightarrow \: b b \ell \nu$ & 
  $3.93\times 10^{-5}$& 1 \\ 
$t(\text{inc.}) + W(\ell\nu)$ & 
  $8.77\times 10^{-5}$ & 16 \\ 
$t(\text{inc.}) + W(\ell\nu) + j $ & 
  $1.54\times 10^{-4}$& 48 \\ 
$W + \bar b b \: \rightarrow \: \bar b b \ell \nu$ & 
  $1.26\times 10^{-5}$ &  2\\ 
$Z + \bar b b \: \rightarrow \: \bar b b \nu \bar{\nu}$ & $5.14\times
   10^{-5}$   & 13 \\
$WZ + {\rm jets}: \rightarrow 3\ell + \nu + {\rm jets}$ & $2.4\times
   10^{-4}$ & 3 \\  
$W + {\rm jets}: \rightarrow \ell\nu+{\rm jets} $ & & \\
$W + j$ & $0.0$ & 0 \\
$W + jj$ & $1.08\times 10^{-5}$ & 80 \\
$W + 3j$ & $3.91\times 10^{-6}$ & 15 \\ 
$Z + \text{jets}\: \rightarrow \: \ell^+\ell^- + \text{ jets}$ & & \\
$Z + j $ & $0.0$ & $0$ \\
$Z + jj $ & $1.55\times 10^{-7}$ & $1$ \\
$Z + \text{ more than } 2j$ & $0.0$ & $0$ \\ 
$Z + \text{jets}\: \rightarrow \: \bar{\nu}~\nu+ \text{ jets}$ & & \\
$Z + j $ & $0.0$ & $0$ \\
$Z + jj $ & $5.2\times 10^{-6}$ & $22$ \\
$Z + \text{ more than } 2j$ & $1.04\times 10^{-6}$ & $2$ \\  \hline
Total Background & & 475 \\\hline  
\end{tabular}
\caption{Cut efficiency and number of post-cut events assuming an
  integrated luminosity of $10\ \fb^{-1}$.} 
\label{table:eff}
\end{table}

\begin{figure}[!ht]
\centering
\begin{minipage}[c]{0.45\linewidth}
   \centering
    \includegraphics[width = 2.5 in, height = 3.5in]{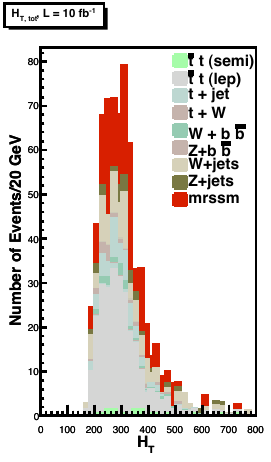} 
    \label{fig:mrssm_ht}
\end{minipage}
\hspace{0.5 cm}
\begin{minipage}[c]{0.45\linewidth}
   \centering
   \includegraphics[width = 2.5 in, height = 3.5in]{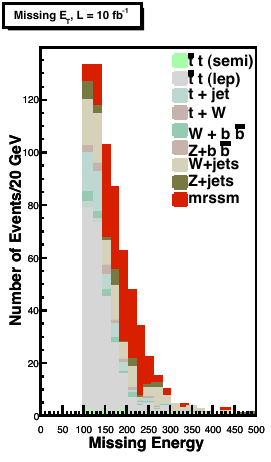}
   \label{fig:mrssm_ETs}
\end{minipage}  
   \caption{Number of events as a function of $H_T$(left figure) and
     $\slashchar{E}_T$ (right figure) assuming an
     integrated luminosity of $10\ \fb^{-1}$ at $14~\tev$.
     The MRSSM signal is for the spectrum given in
     Sec.~\ref{sec:sample}.}
   \label{fig:mrssm_htEt}
\end{figure}

Weighting the generated events by their respective cross sections, we
combined the signal and background histograms for several
variables.  We show the results in Fig.~\ref{fig:mrssm_htEt}.

\subsubsection{Signal Significance as the Spectrum Varies}

\begin{figure}[!ht]
\centering
   \includegraphics[totalheight=0.6\textheight]{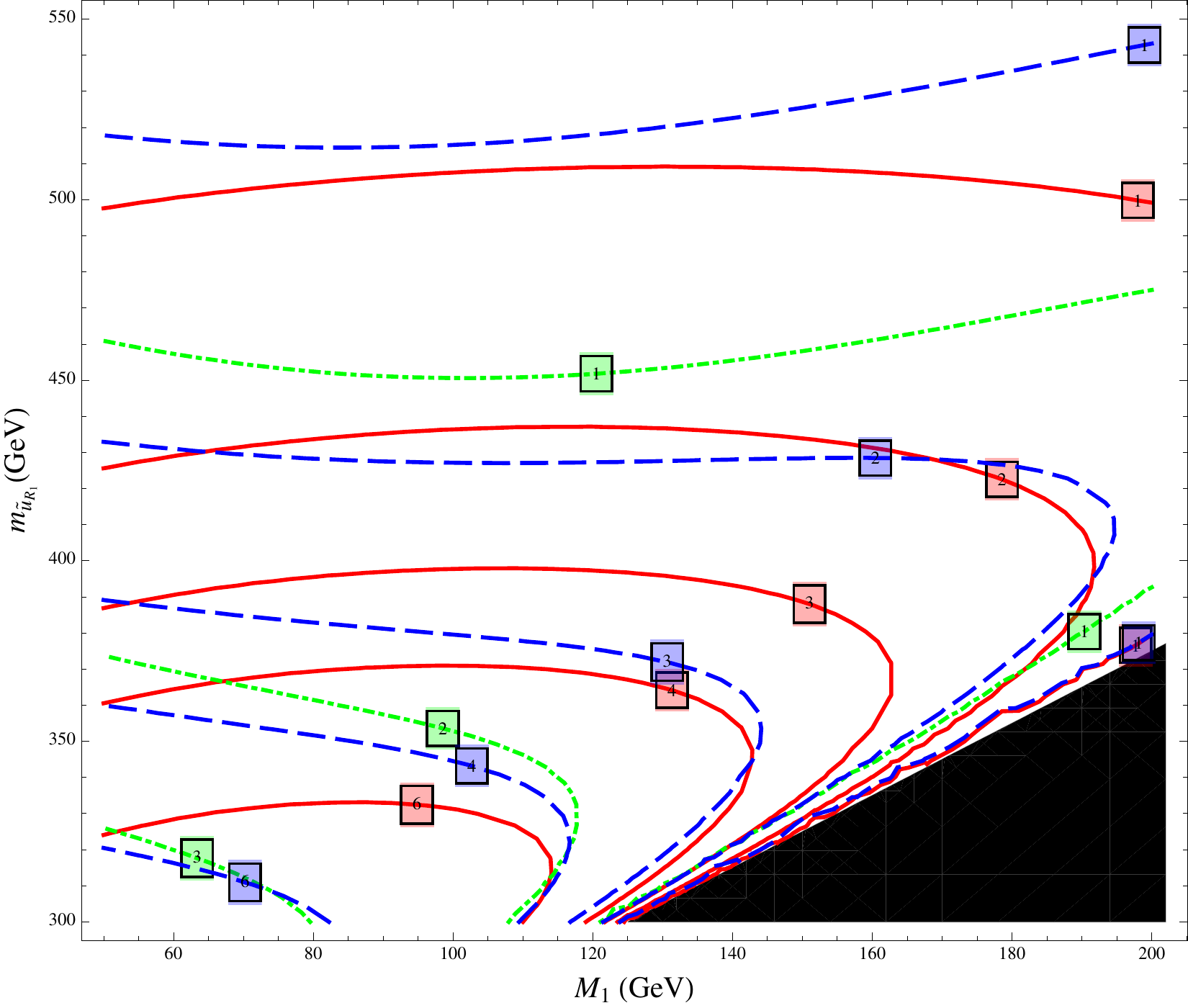} 
   \caption{ Significance $\mathcal{S}$ 
     of the signal  assuming an integrated
     luminosity of $30\ \fb^{-1}$ at $14\ \tev$. In order to produce
     this figure we have assumed the spectrum shown in
     Eq.~\eqref{sig:param}. Red (solid) contours are for $\theta_R =
     \pi/3$, blue (dashed) contours corresponds to  $\theta_R = \pi/4$
     and finally green (dot-dashed) contours are due to  $\theta_R =
     \pi/6$. The channels responsible for background events are listed
     in Table.~\ref{table:sigbr} and we use the same set of cuts as
     described in the subsection.~\ref{singletop-cuts}. In the black
     region on the bottom right corner the neutralino is too heavy and
     a single top event is kinematically  inaccessible. 
   }
    \label{fig:sig}
\end{figure}

Up to now we have identified a set of cuts that separates background 
from signal using a particular sparticle spectrum.  Here we show
that the same set of cuts can also separate signal from background
over a wider range of model parameters.  We characterize our
ability to find the signal over background using ``significance'',
\begin{equation}
\mathcal{S} = \frac{S}{\sqrt{S + B}} \; ,
\end{equation}
where $S$ is the number of signal events and $B$ is the number of
background events.  

Using the same set of cuts listed in Sec.~\ref{singletop-cuts},
we have calculated the significance of the signal spanning
over a grid in $m_{{\tilde{u}_{R_1}}}$ and $M_1$.  
We held the following parameters constant 
\begin{gather}
 m_{\tilde u_{L_1}}  =  1~\tev,\qquad \qquad
 m_{\tilde u_{L_3}} = m_{\tilde u_{R_3}} = 1~\tev, \notag \\
 M_2 = 1~\tev, \qquad M_3 = 3~\tev, \qquad 
\mu_u = \mu_d = 1~\tev,\\
\sin \theta_L = 0 \; \text{ and} 
\quad \tan{\beta} = 10 \; . \notag
\label{sig:param}
\end{gather}
Our result is shown in Fig.~\ref{fig:sig}, where we have interpolated
the grid to form the contours of $\mathcal{S}$ as shown.
Even without optimizing the cuts, we see that the flavor-violating
significance exceeds $1$ for squarks up to about $500$ GeV and
neutralino LSP masses up to about $250$ GeV\@.  Further optimization, 
suitable for specific collider detector characteristics, can surely 
probe even larger regions of parameter space or increase the
significance at any given spectrum point. 

When $\theta_R = \pi/4$ the mass eigenstate squark is an equal mixture
of top and up squarks, so one might expect that the largest single-top
signal comes in this scenario. However, as we can see from
Fig.~\ref{fig:sig} this depends on the mass of the squark. When the
squark mass is comparable to the top quark mass, decays $\tilde u
\rightarrow t + \chi_1$ are kinematically suppressed and the overall
single-top signal suffers. For larger mixing angles (like $\theta_R =
\pi/3$), the mass eigenstate squark is more top-squark than
up-squark. This offsets the kinematic suppression at low squark mass
and leads to a larger single-top signal. 

\subsubsection{Further Refinements}

The analysis in the last section concentrated on eliminating the SM
backgrounds, however there will certainly be additional supersymmetric
contributions to the $b + \ell + j + \slashchar{E}_T$ signal which we
would like to eliminate. These supersymmetric backgrounds are more
difficult to  
gauge without knowing the details of the complete spectrum. 
However, one process which will surely contribute as a background 
is $pp \rightarrow \tilde q \tilde q^* \rightarrow t\bar t$, 
where one of the $b$ quarks goes untagged.  In fact, the correlation 
between the $pp \rightarrow \tilde q \tilde q^* \rightarrow t\bar t$ 
and the $pp \rightarrow \tilde q \tilde q^* \rightarrow t + {\rm \jet}$ 
is crucial to isolate the flavor-violating physics.  
Any enhancement to $t \bar t$, even in the absence
of flavor violation -- such as 
$pp \rightarrow \tilde q \tilde q^* \rightarrow t\bar t$ 
in the flavor-blind MSSM --
will lead to an excess in the final state signal we examined here 
coming from missed or mistagged $b$-jets.  However, 
flavor violation will lead to a \emph{decrease} in the $t\bar t$ signal 
and an \emph{increase} in the single-top signal relative to a flavor-blind
model.  Consider a scenario where the $\tilde q \tilde q^*$ cross
section is $\sigma$, and let us take just one mixing angle and neglect
the mistag rate which affects all channels identically.  Squark pair production
in a flavor-blind model (i.e., top squark production) leads to all events 
in $t \bar t$ and none in single top.  Conversely, squark pair production 
in a model with squark flavor violation will generate 
$\sigma \cos^4{\theta}$ in $t \bar t$ and 
$\sigma \cos^2{\theta}\sin^2{\theta}$ in single top,
assuming the kinematics of the process is held fixed as the 
mixing angle is changed.  In practice, the
comparison is more difficult due to differences in the analysis cuts
applied and multiple squark mixing angles. 

In the SM, an asymmetry in the charge of the final state lepton can 
help improve the single-top significance~\cite{Bowen:2005xq}, 
however this is not the case for the MRSSM flavor-violating signal. 
The lepton asymmetry for SM single-top due to larger $u$, $d$ PDFs 
compared to $\bar{u}, \bar{d}$. The dominant parton-level contribution 
to our signal is $gg \rightarrow \tilde{u}_R \tilde u^*_R$, so we 
are insensitive to the quark PDFs. The final state squark and 
anti-squark are equally likely to decay to a top (anti-top), 
thus we expect an equal number of positively and negatively charged leptons. 
 
There are several ways in which one could improve upon the
analysis presented here.  First there are more sophisticated
optimization techniques (neural net, decision trees, etc.) which can extract 
the correlations between the observables better than we can do by hand.  
Second, we can use the fact that our signal has only one
$b$-quark, while the dominant background (both SM and supersymmetric)
has two.  One way to take 
advantage of this would be to use a $b$-tagging scheme which places 
a premium on getting the $b$'s tagged correctly.  While the analysis
presented here uses the default PGS $b$-tag parametrization, which
has a tag rate of $\sim \! 50\%$ and a mistag rate of $1\!-\!2\%$,
identification rates as high as $90\%$ are possible~\cite{Aad:2009wy}. 
High tag rates have correspondingly high fake rates $\sim \! 30\%$, 
however the backgrounds to our flavor-violating signal which
come from fake $b$-jets are all highly suppressed by the $m_{T,W}$ cut. 
Thus we could hopefully reduce all $t \bar t$ backgrounds
without causing a huge enhancement in backgrounds such as 
$W + {\jets}, W + b\bar b$.  A dedicated study of the viability of this
improved b-tag technique is, however, beyond the scope of this paper.

\section{Case II: A neutralino NLSP and Gravitino LSP}
\label{sec:case2}

We now turn to considering squark flavor violation when the
neutralino decays within the detector and part of the neutralino
energy can be measured by detecting visible particles that result out
of the neutralino decay.  

In this scenario the gravitino is the LSP, which escapes the detector
as missing energy.  As before, we are interested in the 
pair production of squarks which decay to quarks of different flavors 
and the lightest neutralinos.  The new feature is that the 
neutralinos decay to the gravitino plus a photon/$Z$/neutral Higgs. 

\subsection{Neutralino NLSP Decay Models}

The two-body decay width of a neutralino into the   
gravitino and a spectator particle ($X$) is simply given
as~\cite{Ambrosanio:1996jn}:    
\begin{equation}
\label{eq:gravwidth}
\Gamma(\tilde{\chi} \rightarrow \tilde G + X) =
\frac{\kappa~m^5_{\tilde{\chi}}}{96\pi M^2_{\text{pl}}\tilde{m}^2_{3/2}}
\Big( 1 - \frac{m^2_X}{m^2_{\tilde{\chi}}} \Big)^4 \; ,
\end{equation}
where $\kappa$ is an order one mixing angle.  The range of NLSP masses 
that we consider is typically comparable to $Z$ or Higgs mass, 
and thus decays to the $Z$ or $h$ are kinematically suppressed 
compared to the decays to a photon.  For example, when
$\kappa_{Z\tilde G} \approx \kappa_{\gamma\tilde G}$, we find 
$\Gamma_{\gamma \tilde G}/\Gamma_{Z \tilde G}\approx 0.16$ for a 
$150$~GeV NLSP\@.  Three-body decays through an off-shell $Z$ or 
Higgs are suppressed even further.  The only exceptions happen when
$\kappa_{\gamma \tilde G} \ll \kappa_{Z \tilde G}$, which occurs 
when the NLSP is Higgsino-like, i.e., 
$\mu_u \,\text{and/or}\, \mu_d \ll M_1, M_2$.  A Higgsino NLSP,
however, leads to a substantial suppression of the flavor-violating
signal, due to the large top quark Yukawa coupling.  This is clearly
evident by looking at the rightmost figures in Figs.~\ref{fig:1},\ref{fig:2}.
We will see
that this seriously impacts the prospects for finding flavor-violation
with a chargino NLSP in Sec.~\ref{sec:case3}. 
In the case of a Higgsino-like neutralino NLSP, substantial suppression 
is also expected.  It would take a dramatic signal, such as a moderate 
lifetime neutralino NLSP decaying into a $Z$ with 
significant measurable impact parameter, to find evidence of 
squark flavor-violation in this case.

\subsection{Setup and Feasibility}

\begin{figure}[t]
\centering
\includegraphics[width=3in]{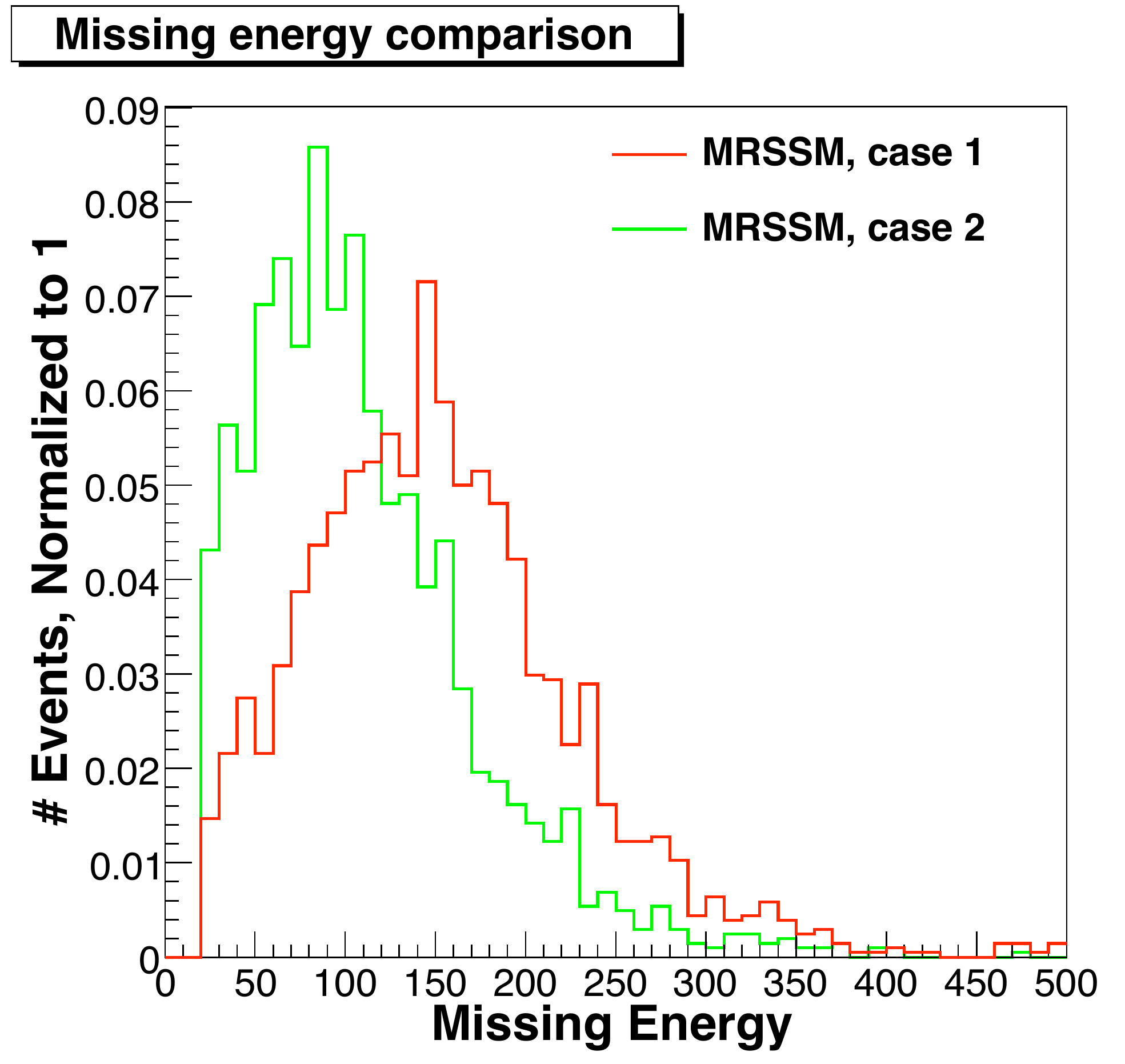}
\caption{Difference in $\slashchar{E}_T$ in MRSSM signals in Case I
  and II.  In Case II, we took $m_{\tilde{G}} = 1$~eV\@.
  For comparison purposes, only soft cuts were imposed:  $\ge
  2\ \jets, p_T >  20\ \gev, \slashchar{E}_T > 20\ \gev, 1\ {\rm
    lepton}, p_T > 25\ \gev$.} 
\label{fig:MET}
\end{figure}

Given the discussion above, the remainder of Case II focuses
on neutralino NLSP decay to a photon.
The flavor-violating signal arises from:
\begin{align}
  \label{eq:process2}
  p + p \ \rightarrow  \  & \left\{ \
   \begin{matrix}  \tilde u_{L_a} + \tilde u_{L_a}^* \\ 
                   \tilde u_{R_a} + \tilde u_{R_a}^* 
   \end{matrix} \right\} \ 
       \rightarrow  \  \text{top + jet} +  \chi_1 
           + \overline  \chi_1  \  \notag \\
      & \rightarrow  \  W + \text{b jet 
          +  jet + 2 photons} 
           + \text{ 2 gravitinos}  \; , \\
  \label{eq:process3}
  p + p \ \rightarrow  \  & \left\{ \
   \begin{matrix}  \tilde d_{L_a} + \tilde d_{L_a}^* \\ 
                   \tilde d_{R_a} + \tilde d_{R_a}^* 
   \end{matrix} \right\} \ 
       \rightarrow  \  \text{b jet + jet} +  \chi_1 
           + \overline  \chi_1  \  \notag \\
      & \rightarrow  \  \text{ b jet 
          +  jet + 2 photons} 
           + \text{2 gravitinos}  \; .
\end{align}
The signal consists of two hard
photons, one $b$ jet and a jet of any other flavor. In the case of up-type
squarks we look for an additional lepton that results from the leptonic
decay of the top.  The neutrino and gravitinos constitute the missing
energy $\slashchar{E}_T$. However, 
as mentioned before, we expect that the missing energy is less whenever
the neutralino decays inside the detector, since part of the neutralino
energy/momentum is redeposited in the detector in the form of
(visible) photons. This can easily be verified by  comparing missing
energy distributions for identical spectrum in Cases I and II as shown
in Fig.~\ref{fig:MET}.

Standard model backgrounds with two hard photons 
are relatively rare, making this scenario of flavor violation nearly
free of SM backgrounds. Exactly what information can be gleaned from
the flavor violating signal depends on the characteristic decay 
length of $\tilde{\chi}_1$:   
\begin{equation}
\label{eq:decay-length}
L = 34\ \kappa^{-1}\ 
\left( E^2/m_{\tilde{\chi}_1}^2 - 1 \right)^{1/2} 
\left( \frac{100~\gev }{m_{\tilde{\chi}_1}} \right)^{5}
\left( \frac{\tilde m_{3/2}}{1~\text{eV}} \right)^2
\ \mu\text{m} \; 
\end{equation}
where $E$ is the energy of the NLSP in the lab frame.

For prompt neutralino decays, the two photons from the processes
shown in Eqs.~(\ref{eq:process2},\ref{eq:process3}) 
point back to the primary vertex. While there is no
displaced vertex, the relatively high transverse momenta ($p_T$) of
these prompt photons is still highly effective for background rejection. 
Note that, since the photons provide the trigger and excellent
background rejection, one can extend this study to find 
flavor-violating squark mixing resulting from the down type squark 
mass matrix, which was not possible in Case I.

If the neutralino is long-lived, while still decaying before it enters
the electromagnetic calorimeter, the signal becomes much more
interesting but also somewhat more complicated.
Because of the finite opening angle between the resulting
photon and the gravitino, the photon may not point back to the primary
interaction point. The identification of a
non-pointing photon has several advantages, since it makes the signal
virtually free of Standard Model background.
However, non-pointing photons may enter the calorimeter at 
a significant angle, causing the electromagnetic shower to be 
spread out over a larger number of cells making the shower shapes 
wider and leading to losses in reconstruction and efficiency.
Detailed search strategies and efficiencies for the non-pointing 
photons in ATLAS can be found in \cite{:1999fr, Prieur:1019876, Aad:2009wy}. 
The electromagnetic calorimeter can be used 
not only to measure the direction and the time of the electromagnetic
shower, but also to determine the mean lifetime of the neutralino.
As long as the neutralino decays before the calorimeter, both $\eta$ and 
$\phi$ can be measured and a vector corresponding to the path of the
photon can be constructed. Although the exact decay point cannot be
measured, the path of the photon can be extrapolated to the beam
axis, and further information can be extracted from the the distance
between this point and the primary vertex. Also photons produced from
long-lived neutralinos arrive later than the ones directly from the
primary vertex.  Calorimeter timing information can be used to
gather further information about the lifetime of the neutralino.    

If neutralinos are sufficiently long-lived so that they decay
well outside the detector, the resulting photons will be missed. 
This ``collider-equivalent'' LSP leads to a collider signal that is 
exactly the same as that studied in Case I (see Sec.~\ref{sec:case1}).

\subsection{Sample Signal with Background}

Having identified the different scenarios within Case II, we now
examine the signals and background for a simple subset of the parameter
space. 
We use the same spectrum parameters as in Case I:  all superpartners
are heavy $(\ge 1\ \tev)$ except for one neutralino with mass
$m_{\tilde{\chi}_1} = 50\ \gev - 200\ \gev$, and one right handed
up-type squark, $m_{\tilde u_R} = 300 - 500\ \gev$. The spectrum
parameters must be supplemented with a gravitino 
mass which we take to be $1$~eV\@.  We are free to make this
choice independent of the other soft masses, and it leads to a
neutralino lifetime of order few nanoseconds and thus a prompt
decay. The restriction to prompt decays is necessary as our software
tools are inadequate to accurately depict scenarios with displaced
vertices. The signal events we look for consist of $\ell + b + \jet +
\gamma \gamma + \slashchar{E}_T$.  

\subsubsection{Backgrounds}

The primary SM backgrounds for this case are the same as in Case I,
however we are now interested in the fraction of events
which contain final state photons. Photons appear in the final state
of these backgrounds due to bremsstrahlung, from hadron decay
products, or from jets faking photons.   The extra photons require a
price of 
$\alpha_{em}$ if they are real, or a factor of the jet-photon fake
rate if the photons are actually jets faking photons. Real photons emitted as
  bremsstrahlung are 
usually soft,  while photons from subsequent hadron decays are usually
buried within a jet. 
 The jet-photon
fake rate is $p_T$ and $|\eta|$ dependent 
  and is expected to be $\sim 0.1\! -\! 0.01\%$~\cite{:1999fr,
    Ball:2007zza}. Furthermore, the $p_T$ spectrum of the fake jets depends on how one models faking.  Since the fake rate is model dependent and sufficiently less than $\alpha_{em}$, we will not consider faked photons in this work . 
%

The signal, by contrast, results from a neutralino decaying to a 
gravitino plus photon. For practical purposes, the branching ratio is $\mathcal O(1)$.
Moreover, the photons from the signal can be hard,
$p_{T,\gamma} \sim m_{\chi_1}/2$. Thus, by simply requiring the
presence of final state photons, we expect all SM backgrounds can be
substantially reduced compared to their 
counterparts in Case I\@.

\subsubsection{Cuts}

\begin{table}[t]
\centering 
\begin{tabular}{ | c | c | }   \hline 
Process & \# events in $10\ \fb^{-1}$  \\ 
\hline \hline &  \\
$\tilde u_{R,i} \tilde u^*_{R,i} \: \rightarrow \: t j \chi_1 \,
  \overline{\chi}_1 \: \rightarrow \: \ell b j \gamma\gamma +
  \slashchar{E}_T$ & 481 \\   
 ~ $m_{{\tilde u}_R} = 300\ \gev, ~ m_{\tilde{\chi}_1} = 50\ \gev$ & \\
  & \\   \hline \hline
$t \bar t \: \rightarrow \: b\bar b j \ell \nu $ & 1.3 \\ 
$t\bar t \: \rightarrow \: \ell \ell' \nu \nu'$ & 1.4 \\ 
$t + q \: \rightarrow \: b j \ell \nu$ & 0 \\ 
$t + b \: \rightarrow \: b b \ell \nu$ & 0 \\ 
$t(\text{inc.}) + W(\ell\nu)$ & 0 \\ 
$t(\text{inc.}) + W(\ell\nu) + j $ & $\le 1$ \\ 
$W + \bar b b \: \rightarrow \: \bar b b \ell \nu$  & 0\\ 
$Z + \bar b b \: \rightarrow \: \bar b b \nu \bar{\nu}$ & 0 \\
$WZ + {\rm jets}: \rightarrow 3\ell + \nu + {\rm jets}$ & 0 \\  
$W (\ell \nu) + {\rm jets}$ & $\le 1$ \\
$Z(\ell^+\ell^-) + \text{jets} $ & 0 \\
$Z(\bar{\nu}\nu) +\text{jets} $ & 0 \\ \hline 
Total Background & $\le 5$ \\ \hline  
\end{tabular}
\caption{Number of post-cut events assuming an
  integrated luminosity of $10\ \fb^{-1}$. The same cross sections and
  event samples were used as in Sec.~\ref{sec:case1}. Zero events indicates 
  no events in the generated data sample passed the cuts, while $\le 1$
  indicates some events did pass the cuts, but, once properly
  normalized, the number of events was some fraction less than
  $0.5$. The $W/Z + \text{jets}$ entries are the sum of the of the $+j, +2j, \text{ and}+\text{more than } 2j$ backgrounds. Because the cut on the transverse mass of the $W$ is no
  longer needed to suppress SM background, the number of signal events
  in this case is substantially larger than the corresponding point in
  Case I\@.} 
  \label{table:case2eff}
\end{table}

\begin{figure}[!ht]
\centering
\includegraphics[totalheight=0.35\textheight]{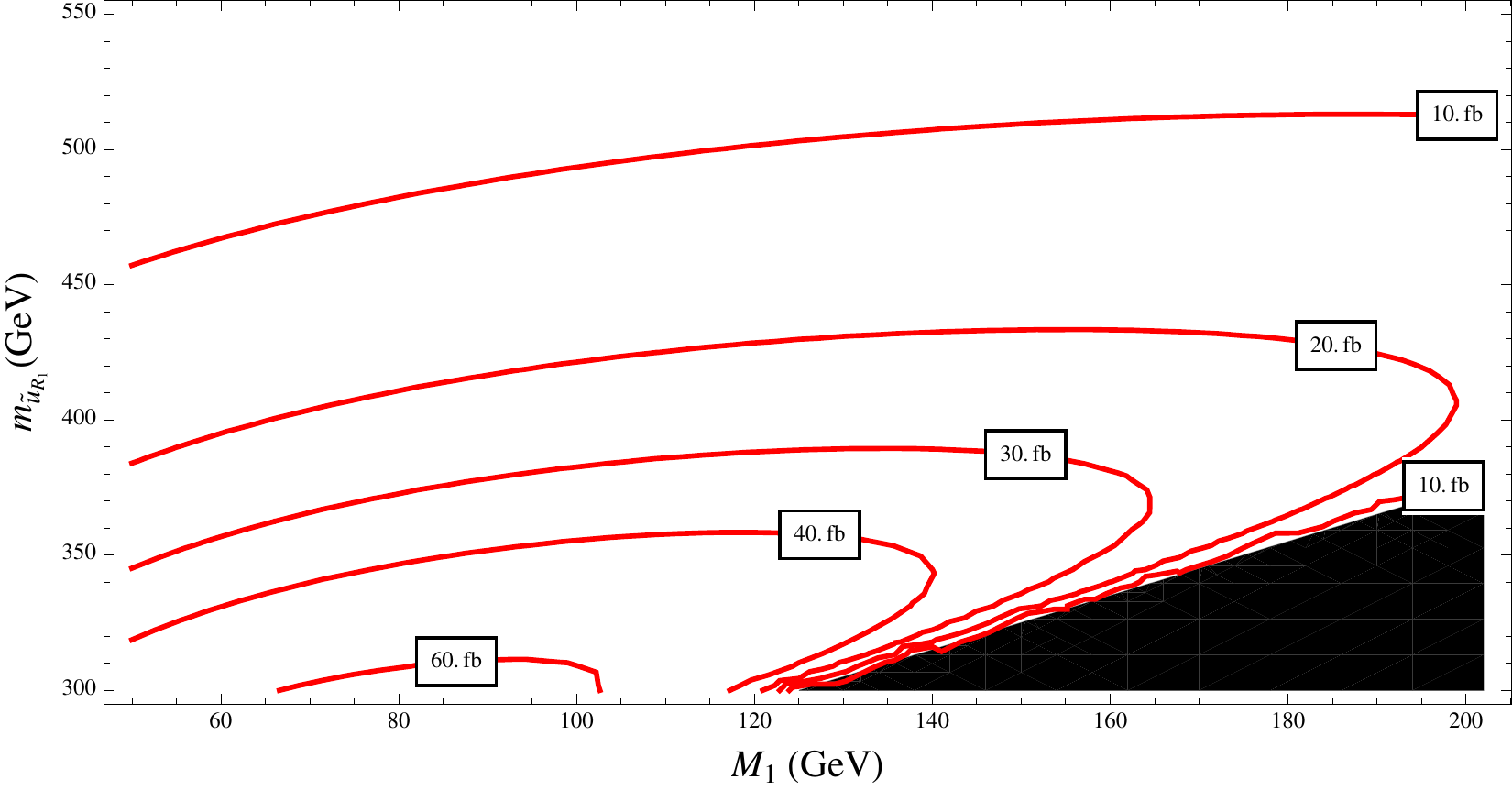}
\caption{Signal cross section as a function of $\tilde q_R$ mass and
  $\chi_1$ mass, assuming $\theta_R = \pi/3$. As before, the area in the
  bottom right is kinematically forbidden.} 
\label{fig:case2sig}
\end{figure}

From the above argument, we expect the combination of hard photons
with typical SUSY features (large $\slashchar{E}_T$, many high $p_T$
objects) has little SM background. This was verified in a recent ATLAS
study of GMSB scenarios~\cite{Aad:2009wy}. Specifically, in
Ref.~\cite{Aad:2009wy} the 
$W/Z +\ \jets\ \text{ and } t \bar t + \jets$ 
backgrounds were essentially eliminated by requiring two hard ($p_T >
20~\gev$) photons on top of conventional SUSY cuts ($\ge 4\ \jets,
p_T > 50~\gev$, leading jet $p_T > 100~\gev$, $\slashchar{E}_T >
100\ \gev$, $\slashchar{E}_T > 0.2\times M_{eff}$). However, the
ATLAS study only looked at discovery potential for benchmark GMSB
points, while we are looking to pick out a flavor violation signal and
therefore we need to tailor the cuts beyond what was done in 
Ref.~\cite{Aad:2009wy}. First, we require a single b-tag and a
lepton. Second, because 
we considered the shortest possible signal decay chains, the number of
jets in the signal after showering and hadronization is less than in a
typical GMSB benchmark point. To account for this we reduce the
minimum number of jets required in the signal to $2$.  Since we have
adjusted the cuts, and are looking for events with tagged b-jets, we
have included the backgrounds $W + \bar b b + \jets, Z + \bar b b +
\jets, t + \{q, b, W \} + \jets$ as well as the backgrounds studied in
Ref.~\cite{Aad:2009wy}. The final cuts we impose in Case II are thus:  
\begin{itemize}
\item Exactly two photons, $p_T > 20~\gev, |\eta| < 2.5$. These
  photons serve as our primary trigger. 
\item Exactly one lepton ($\ell = e, \mu)$ with $p_T > 30~\gev, |\eta| < 2.5$.
\item $\ge 2$ jets, $p_T > 50~\gev$. The leading jet must have $p_T >
  100~\gev$ and exactly one jet must be tagged as a b-jet. 
\item $\slashchar{E}_T > 100~\gev$.
\item $\slashchar{E}_T > 0.2\times M_{eff}$, where $M_{eff} = \sum_{i
    = j, \ell, \gamma}p_{T,i} + \slashchar{E}_T$. 
\end{itemize}

Although we have changed the cuts from Ref.~\cite{Aad:2009wy}, we reach the
same conclusion: the requirement of two hard photons renders all 
SM backgrounds to be negligible. To be explicit, the number of 
backgrounds events,
assuming $10~\fb^{-1}$ of integrated luminosity and after imposing
the above cuts is shown in Table~\ref{table:case2eff}. 
A sample signal point, with 
$m_{{\tilde u}_R} = 300~\gev, ~ m_{\tilde{\chi}_1} = 50~\gev$, 
is also shown.  

With virtually no SM backgrounds, a flavor violation signal 
might be extracted from just a handful of events. Of course, as in Case I, 
some supersymmetric backgrounds will survive the cuts above --
but these are difficult to quantify. To demonstrate the order of
magnitude signals one can expect in this scenario, we therefore show
in Fig.~\ref{fig:case2sig} the MRSSM signal cross section after 
imposing the cuts above in a sample region of parameter space.  
Hundreds of background-free flavor-violating events are expected 
in the region shown given just $10$~fb$^{-1}$ of data.

\section{Case III: A Chargino NLSP and Gravitino LSP}
\label{sec:case3}

Finally we discuss the scenario with a chargino NLSP and a gravitino
LSP\@.  Such a scenario does not ordinarily occur in the MSSM\@. In
the MRSSM it is quite generic \cite{Kribs:2008hq}. 
Given the fact that we only consider large $M_2$, the lightest 
chargino lighter than a neutralino can only occur when one or both
of $\mu_u,\mu_d$ is much smaller than $M_1$.  That necessarily 
implies that the NLSP is mostly one of the charged Higgsinos. 

Just like in the last section, the decay length of the chargino is
crucial.  Since LEP II has ruled out charginos lighter than about
100 GeV~\cite{Amsler:2008zzb}, the 2-body decay $\chi_1^\pm \ra W^\pm \tilde{G}$ is open,
but the width is suppressed by the small wino content of the chargino.
The 2-body decay $\chi_1^\pm \ra H^\pm \tilde{G}$ may also be open,
or, if the charged Higgs is heavier than the lightest chargino,
the decay may proceed through a 3-body decay with an off-shell $H^\pm$.
Exactly which decay dominates is parameter-dependent.
Decays to a $W$ have the advantage of the nontrivial branching
fraction of the $W$ into $e$ and $\mu$.  The decay into or through
a charged Higgs $H^+$ could result in $t \, \bar b$ (if open), or 
$\tau^+ \, \nu_\tau$ or $c \bar s$~\cite{Amsler:2008zzb}.  The leptonic branching
fraction of a chargino NLSP, therefore, may be rather suppressed.

A larger problem lies in the intrinsic suppression of the flavor-violating
decays.  Higgsinos mostly couple to the third generation quarks.
Even when the mixing angle in the squark mass matrix is maximal,
the large third generation Yukawa couplings and small gaugino content 
imply that the squarks overwhelmingly decay to the quarks of 
third generation.  This makes it extremely challenging to generate
a flavor-violating signal through squark decay.

If the decay length of the chargino (in this case a Higgsino) is
large enough so that isolated charged tracks of the chargino NLSPs 
are visible, this dramatic signal may readily be found.
Certainly a clearly visible isolated track will significantly
reduce (or potentially eliminate) standard model 
backgrounds~\cite{Nisati:1997gb,Allanach:2001sd,Prieur:1019876,Ellis:1006573,Giagu:2008im}.
Below we list the decay chain we look for (chargino decays are not shown): 
\begin{align}
  \label{eq:process4}
  p + p \ \rightarrow  \  & \left\{ \
   \begin{matrix}  \tilde u_{L_a} + \tilde u_{L_a}^* \\ 
                   \tilde u_{R_a} + \tilde u_{R_a}^* 
   \end{matrix} \right\} \ 
       \rightarrow  \  \text{b jet + jet} +  \chi_1^{+} 
           + \overline{\,\chi_1^{+}}  
             \; ,  \\ 
  \label{eq:process5}
  p + p \ \rightarrow  \  & \left\{ \
   \begin{matrix}  \tilde d_{L_a} + \tilde d_{L_a}^* \\ 
                   \tilde d_{R_a} + \tilde d_{R_a}^* 
   \end{matrix} \right\} \ 
       \rightarrow  \  \text{top + jet} +  \chi_1^{+} 
           + \overline{\,  \chi_1^{+}}  \  \notag \\
      & \rightarrow  \  W + \text{b jet+  jet } +  \chi_1^{+} 
           + \overline{\,\chi_1^{+}}  \; .
\end{align}
The event then consists of two isolated charged tracks, 
two jets with one of them tagged as a $b$-jet, and no
missing $\slashchar{E}_T$\@. As shown in Eqs.~\eqref{eq:process4} and
\eqref{eq:process5} additional leptons would be present from $W$ decay
when the pair-produced squarks are of down-type flavor. 
The effective mass of all final state particles can be 
reconstructed to estimate the squark masses. 

\section{Conclusions}
\label{sec:conclusions}

We have studied the production of the single top events at the LHC as 
a signal of flavor violation in the squark mass matrices of the MRSSM. 
The results may be summarized as follows:
\begin{itemize}
\item  We considered the shortest decay chain where pair-produced 
squarks decay to quarks of different flavors and gauginos. 
The gaugino content of the lightest neutralino or chargino
is either bino-like or Higgsino-like, given the constraint
on $M_2$ \cite{Kribs:2007ac}.  Since the squark-Higgsino-quark 
coupling dominantly involves the third generation quarks, 
pair-produced squarks decay to pairs of tops whenever the lightest neutralino is mostly a Higgsino.  As a result, 
the flavor-violating single top events are highly suppressed 
with respect to the double top events.  Hence, the best case scenario 
to produce single top events is when the bino is the lightest neutralino.

\item We performed a thorough analysis of the single top signal 
and background, when both neutralinos escape the detector. 
This is the so-called ``collider equivalent LSP'' Case I.
We devise a set of cuts that reduce the background, achieving 
large ``significance'' of the single top signal.  Furthermore, 
we estimated the significance of the signal as the squark mass
and neutralino mass were varied.

\item The best case scenario, our Case II, to detect flavor violation
is when the neutralino decays within the detector to the gravitino 
and a photon.  The signal contains two hard photons (generated at 
the interaction point or from secondary vertices, depending on the 
decay length of the neutralino) and provides virtually background 
free events.  We explicitly showed how the size of the signal 
varies as the squark and neutralino masses were varied.

\item  In $R$-symmetric models, a chargino 
is often lighter than the lightest neutralino \cite{Kribs:2008hq}.
If the chargino is long-lived, it produces a charged track of a 
heavy object that will provide a nice handle to significantly 
reduce background events.  While the production rate of flavor-violating 
signals suffers (the squark-Higgsino-quark coupling is dominated
by the third generation)the unusual character of a long-lived
chargino may well provide a means to identify a flavor-violating
signal.


\end{itemize}

\section*{Acknowledgments}

We thank Tim Tait and Carlos Wagner for very useful comments.
GDK and AM thank the Aspen Center for Physics for
hospitality where part of this work was completed.
This work was supported in part by the DOE under contracts
DE-FG02-96ER40969 (GDK, TSR) and DE-FG02-92ER40704 (AM).

\begin{appendix}

\renewcommand{\theequation}{A-\arabic{equation}}
\setcounter{equation}{0} 
\renewcommand{\thetable}{A.\arabic{table}}
\setcounter{table}{0}

\section{Appendix:  An $R$-symmetric Model, Mass Matrices, and the \\ 
Interaction Lagrangian}

For the purpose of model building and calculations of various observables, 
the minimal $R$-symmetric supersymmetric model (MRSSM) is usually
written in terms of two component Weyl fermions - in terms of which
physics is simple and  transparent. We however are interested in the
collider phenomenology. Using modern Monte Carlo tools necessitates the 
construction of a formalism where all the fermions are in  four
component notation and all particles are represented by mass eigenstate fields.
In this section we provide the notation and the details of one 
such formalism suitable for implementation into MadGraph.
This section to be self-sufficient.  Also, we only provide details 
of all the interactions that are of relevance for this paper 
(namely, fermion+sfermion+gaugino interactions).

\subsection{Particle Content}

First we list all the particles along with their quantum
numbers under the standard model gauge group:  
\begin{table}[h]
\centering 
\begin{tabular}{c|cccc}
  Fields \quad & $SU(3)_C$ & $SU(2)_W$ & $U(1)_Y$       & $U(1)_R$ \\ \hline 
  $Q$          & $3$       & $2$       & $\frac{1}{6}$  & $1$   \\
  $U$          & $\bar{3}$ & $1$       & -$\frac{2}{3}$ & $1$   \\
  $D$          & $\bar{3}$ & $1$       & $\frac{1}{3}$  & $1$   \\
  $L$          & $1$       & $2$       & -$\frac{1}{2}$ & $1$   \\
  $E$          & $1$       & $1$       & $1$            & $1$   \\
  $\Phi_{B}$   & $1$       & $1$       & $0$            & $0$   \\
  $\Phi_{W}$   & $1$       & $3$       & $0$            & $0$   \\
  $\Phi_{g}$   & $8$       & $1$       & $0$            & $0$   \\
  $H_u$        & $1$       & $2$       & $\frac{1}{2}$  & $0$   \\
  $H_d$        & $1$       & $2$       & -$\frac{1}{2}$ & $0$   \\
  $R_u$        & $1$       & $2$       & -$\frac{1}{2}$ & $2$   \\
  $R_d$        & $1$       & $2$       & $\frac{1}{2}$  & $2$   \\
\hline
\end{tabular}
\caption{Gauge and $R$-charges of all chiral supermultiplets in the MRSSM\@.} 
\label{table:fields}
\end{table}

\noindent Table~\ref{table:fields} only contains the matter and Higgs 
superfields. In addition, the MRSSM also contains three
real superfields $B, W$ and $G$ which are in the adjoint
representations of corresponding gauge groups ($U(1)_Y, SU(2)_W,
SU(3)_C$ respectively) and carry zero $R$-charge. 

Next we look in the details of these multiplets. We establish
notation for the components of the superfields, and write the mass
matrices and interactions of interest.
  
\subsubsection{Gauge and Higgs Superfields}
\label{app:gaugehiggs}

The scalars and fermions in the Higgs superfields are denoted as 
\begin{gather}
  \label{eq:higgs}
  H_u =          \begin{pmatrix}
        \{ \tilde H_u^{+}, \;  h_u^{+}  \}  \\ 
        \{ \tilde H_u^{0}, \; h_u^{0} \}
                 \end{pmatrix} , \qquad  
  H_d =          \begin{pmatrix}
        \{ \tilde H_d^{0}, \;  h_d^{0}  \}  \\ 
        \{ \tilde H_d^{-}, \; h_d^{-} \}
                 \end{pmatrix}    \; ,  \\
  R_u =          \begin{pmatrix}
        \{ \tilde R_u^{0}, \;  r_u^{0}  \}  \\ 
        \{ \tilde R_u^{-}, \;  r_u^{-} \}
                 \end{pmatrix} , \qquad  
  R_d =          \begin{pmatrix}
        \{ \tilde R_d^{+}, \;  r_d^{+}  \}  \\ 
        \{ \tilde R_d^{0}, \;  r_d^{0} \}
                 \end{pmatrix}    \; .
\end{gather}
The $R-$Higgs superfields are introduced to generate Higgsino masses
in a $R-$symmetric way. 
The real superfields for the three gauge groups contain the usual spin-$1$ 
gauge fields and their corresponding gauginos: 
\begin{gather}
  \label{eq:gauge}
  B   \rightarrow  \{ B_\mu, \; \tilde{B} \}, \qquad 
  W^a \rightarrow  \{ W_\mu^a, \; \tilde{W}^a \},  \qquad
  G^a \rightarrow  \{ G_\mu^a, \; \tilde{G}^a \}, \\
  \Phi_{B}  \rightarrow  \{ \psi_B, \; \phi_B \}, \qquad 
  \Phi_{W}^a  \rightarrow  \{ \psi_W^a, \; \phi_W^a \}, \qquad 
  \Phi_{G}^a  \rightarrow  \{ \psi_G^a, \; \phi_G^a \}
\end{gather}
All fermions shown in Eqs.~\eqref{eq:higgs} and 
\eqref{eq:gauge} are Weyl fermions in the
$(\frac{1}{2}, 0)$ representation of the Lorentz group. Combining the
two-component spinors into four-component Dirac spinors, we have  
\begin{gather}
\lambda_G^a = \begin{pmatrix} 
             \psi_G^a \\ \tilde{\ G^a}^\dag 
            \end{pmatrix} 
\label{gluino-def}\\
\lambda_B = \begin{pmatrix} 
             \psi_B \\ \tilde{ B}^\dag 
            \end{pmatrix}, \qquad 
\lambda_{3} = \begin{pmatrix} 
             \psi_W^0 \\ \tilde{\ W^0}^\dag 
            \end{pmatrix}, \qquad 
\lambda_{H_1} = \begin{pmatrix} 
             \tilde H_u^0 \\ \tilde{\ R_u^{\, 0}}^\dag 
                \end{pmatrix}, \qquad 
\lambda_{H_2} = \begin{pmatrix}  
             \tilde H_d^0 \\ \tilde{\ R_d^{\, 0}}^\dag 
                 \end{pmatrix}  
\label{nut-gauge-def}  \\ 
\lambda^{+}_{1}  =  \begin{pmatrix}  
             \psi_W^{+} \\ \tilde{\ \: W^{-}}^\dag 
                    \end{pmatrix}, \qquad 
\lambda^{-}_{2} =   \begin{pmatrix}  
             \psi_W^{-} \\  \tilde{\ \: W^{+}}^\dag
                    \end{pmatrix}, \qquad
\lambda_{H_{1}}^{+}  = \begin{pmatrix}  
             \tilde H_u^{+} \\ \tilde{\ R_u^{-}}^\dag 
             \end{pmatrix}, \qquad
\lambda_{H_{2}}^{-} = \begin{pmatrix}  
             \tilde H_d^{-} \\ \tilde{\ R_d^{+}}^\dag 
           \end{pmatrix} \; .
\label{charge-gauge-def}
\end{gather}
Note we have defined the four component spinors to be eigenstates of
$R$-symmetry. In particular, all the four 
component fermions shown in Eqs.~\eqref{gluino-def},
\eqref{nut-gauge-def} and  \eqref{charge-gauge-def}, which are also
gauge eigenstates, have $R$-charge  $-1$.  
The charge conjugates of all the Dirac spinors above are  defined in
the usual way, 
\begin{equation}
  \lambda^c = i \gamma_2 \lambda^*  \qquad \text{and} \qquad
  \lambda^{-} = i \gamma_2 \left( \lambda^{+}\right)^* \; .
\end{equation}

Next we list all operators which generate gaugino and the Higgsino
mass terms: 
\begin{multline}
\int \!\! d^2\theta \ \frac{W'_\alpha}{M} W^{\alpha}_i \Phi_i  + 
 \mu_u H_u R_u + \mu_d H_d R_d   \\
  + H_u \left( \lambda_u \Phi_{W} + 
        \lambda_u' \Phi_{B} \right) R_u +
             H_d \left( \lambda_d \Phi_{W} 
                + \lambda_d' \Phi_{B} \right) R_d \; .
\label{dirac-eq}
\end{multline}
where $W'_\alpha$ is the field strength chiral superfield for an extra
$\text{U}(1)$ and $i$ runs over the three standard model gauge groups
$B, W$ and $G$\@. Gauginos and Higgsinos obtain their masses after
$W'_\alpha$ is expanded around its supersymmetry breaking VEV
and $H_u$ and $H_d$ are expanded around their electroweak symmetry
breaking VEV\@. Note that these are the only operators, involving the
$R$~fields, that are allowed under all symmetries.  

In our gauge eigenstate basis the neutralino masses may now be written
as  
\begin{gather}
  \label{mass-nut}
    \overline{N} \, M_N \,  P_L \,  N  \: + \: \text{c.c.}   \\
    = \begin{bmatrix}
      \bar \lambda_B & \bar \lambda_3 & 
            \bar \lambda_{H_1} & \bar \lambda_{H_2}
    \end{bmatrix}  
    \begin{bmatrix}
      M_1 &  0  & g'v_u/\sqrt{2} & - g'v_d/\sqrt{2} \\
       0  & M_2 & -gv_u/\sqrt{2} &   gv_d/\sqrt{2} \\
\lambda_u' v_u/\sqrt{2} & -\lambda_u v_u/\sqrt{2} & \mu_u & 0 \\   
      -\lambda_d' v_d/\sqrt{2} & \lambda_d v_d/\sqrt{2} & 0 & \mu_d 
    \end{bmatrix} P_L 
    \begin{bmatrix}
    \lambda_B \\ \lambda_3 \\ \lambda_{H_1} \\ \lambda_{H_2}  
    \end{bmatrix} \: + \: \text{c.c.} \; ,
    \notag
\end{gather}
where $P_L$ and $P_R$ are the projection matrices on the $4$ component
Dirac spinors and projects out the top and the bottom Weyl spinors
respectively.
Although the vector $N$ of fermions and the
mass matrix in Eq.~\eqref{mass-nut}  looks similar to the neutralino
mass term in the MSSM, they are drastically different. The vector $N$
is made of $4$ component spinors, which are eigenstates of
$R$-symmetry with $R$-charge $-1$.  
The fact that the neutralinos carry a conserved $U(1)$ charge
($R$-symmetry in this case) makes them pure Dirac spinors.    

Because of the Dirac nature of the neutralinos, their masses are
determined by a biunitary diagonalization of the mass matrix 
$M_N$. This is entirely analogous to the chargino mass matrix.
The squares of the neutralino masses are determined from diagonalizing 
the matrix $M_N^\dag M_N$. In short, in the mass basis, 
\begin{gather}
  \label{eq:mass-nut-estate}
    \overline{\: \chi} \, M_{\chi} \, P_L \, \chi  \: + \: \text{c.c.}  \\
  \chi = \left( L_{N} P_L + R_N  P_R \right) N  
     \quad \text{ and } \quad
  M_{\chi} =   L_{N} M_N R_{N}^\dag     \; ,
\end{gather}   
where $\chi$ are mass eigenstates and $M_{\chi}$ is a diagonal
matrix with the diagonal entrees sorted. The unitary rotations
$L_{N}$ and $R_N$ diagonalize and 
then sort $M_N M_N^\dag$ and $M_N^\dag M_N$ respectively.  
$P_L$ and $P_R$ are the projection matrices on the $4$ component
Dirac spinors and projects out the top and the bottom Weyl spinors
respectively.   
 
Similarly, the charginos of the MRSSM are written in terms of 
the gauge eigenstates as: 
\begin{gather}
  \label{mass-charge}
    \overline{\: C^{+}} \,  M_C  \, P_L \, C^{+}  \: + \: \text{c.c.}  \\
   = \begin{bmatrix}
      \bar \lambda_{1}^{+} & \bar \lambda_{H_1}^{+} & 
            \bar \lambda_{2}^{+} & \bar \lambda_{H_2}^{+}
     \end{bmatrix}  
    \begin{bmatrix}
      M_2 &  gv_u  & 0 & 0 \\
      \lambda_u v_u  & \mu_u & 0 & 0 \\
      0  &  0 &  M_2 &  \lambda_d v_d \\
      0  &  0 &  g  v_d &  \mu_d 
    \end{bmatrix} P_L
    \begin{bmatrix}
    \lambda_{1}^{+} \\ \lambda_{H_1}^{+} \\ 
         \lambda_{2}^{+} \\ \lambda_{H_2}^{+}  
    \end{bmatrix}  \: + \: \text{c.c.} \; .
    \notag
\end{gather} 
Just like the neutralinos, Eq.~\eqref{mass-charge} can be written in
terms of the mass eigenstates
\begin{gather}
  \label{eq:mass-charge-estate}
  \overline{\: \chi^+}  \,  M_{\chi^+} P_L \, \chi^+   \: + \: \text{c.c.}\\
  \chi^{+} = \left( L_{C} P_L + R_{C}  P_R \right) C^{+} 
     \quad \text{ and } \quad
  M_{\chi^+} =   L_{C} M_C R_{C}^\dag   \; ,
\end{gather}

\subsubsection{Matter Superfields}
\label{app:matter}

We take the matter superfields to be:
\begin{gather}
  \label{eq:matter}
    Q \rightarrow \begin{pmatrix} 
               \{ \xi_u, \;  \tilde u_L  \}  \\ 
               \{ \xi_d, \; \tilde d_L \}
                  \end{pmatrix} , \qquad
    U \rightarrow  \{  \xi_{\bar u}, \;  \tilde u^{*}_R \}  , \qquad
    D \rightarrow  \{  \xi_{\bar d}, \;  \tilde d^{*}_R \} 
    \nonumber \\
    L \rightarrow  \begin{pmatrix}  
               \{ \xi_\nu, \;  \tilde \nu_L  \}  \\ 
               \{ \xi_e, \; \tilde e_L \}  
                  \end{pmatrix} , \qquad 
    E \rightarrow  \{  \xi_{\bar e}, \;  \tilde e^{*}_R \} 
\end{gather}
All $\xi_f$ are Weyl fermions in the $(\frac{1}{2}, 0)$ representation 
of the Lorentz group, while $\tilde u_{\lambda}, \tilde d_{\lambda}$,
etc. are the superpartner scalars. We have defined the superfields $U,
D$ and $E$ in such a way that both the left and the right scalar
superpartners have the same charge.

For the purpose of phenomenology it is more convenient to use a notation
where all the fermions are represented in a four component notation.
\begin{equation}
u_i =         \begin{pmatrix} 
           \xi_{u_i} \\ \bar{\xi}_{\bar{u}_i} 
              \end{pmatrix}, \quad  
d_i =         \begin{pmatrix} 
           \xi_{d_i} \\ \bar{\xi}_{\bar{d}_i} 
              \end{pmatrix}, \quad 
l_i =         \begin{pmatrix} 
           \xi_{\ell_i} \\ \bar{\xi}_{\bar{l}_i} 
              \end{pmatrix}, \quad 
\nu_i =       \begin{pmatrix} 
           \xi_{\nu_i} \\ \bar{\xi}_{\bar{\nu}_i} 
              \end{pmatrix} \; ,
\end{equation}
where $i$ is the generation index. Without loss of generality we will
work in a basis where the Yukawa couplings are diagonal (the mass
eigenbasis). 
For example, $\xi_{u_i}$ and $\xi_{\bar u_i}$ correspond to the
$i$th eigenstate of the up-type quark mass terms.

This is, however, \emph{not} the mass basis of the scalar superpartners.
Assuming that only the third generation fermions are massive, we find
the mass terms of the matter superparticles are given by: 
\begin{equation}
  \label{mass-sparticle}
  m_{\tilde f_{ij}}^2 \tilde f_i^* \tilde f_j   = 
  \left( \tilde m_{f_{i j}}^2 
       + m_f^2 \:  \delta_{i 3} \: \delta_{j 3} \right) 
   \tilde f_{i}^*  \tilde f_{j} \; ,
\end{equation}
where $\tilde f$ runs over the species $\tilde u_L, \tilde u_R, \tilde
d_L, \tilde d_R, \tilde e_L, \tilde e_R$ and $\tilde \nu_L$. In
particular, mass parameters given in Eq.~\eqref{mass-sparticle} are 
\begin{equation*}
  \begin{split}
     \tilde m_{u_L}^2 = \tilde m_{d_L}^2 = \tilde m_{Q}^2, \qquad 
     \tilde m_{u_R}^2 = \tilde m_{U}^2, \qquad 
     \tilde m_{d_R}^2 = \tilde m_{D}^2, \qquad 
     \tilde m_{e_L}^2 = \tilde m_{\nu_L}^2 = \tilde m_{L}^2, \qquad
     \tilde m_{e_R}^2 = \tilde m_{E}^2  \\ 
     m_{u_L} = m_{u_R} = m_t, \qquad 
     m_{d_L} = m_{d_R} = m_b, \qquad
     m_{e_L} = m_{e_R} = m_\tau, \qquad  m_{\nu_L} = 0 \; ,
  \end{split}
\end{equation*}
where we have taken the first and second generation SM quark/lepton
masses to vanish.
The physical masses $m_{\tilde f}$ are now diagonalized by unitary
transformations. Each species  $\tilde f \supset \tilde u_L, \tilde
d_L, \tilde u_R, \tilde d_R, \tilde e_L, \tilde e_R$, and
$\tilde{\nu}_L$ has its own transformation, which we designate as $U_{
  f}$. For 
convenience we designate scalars (sfermions) in  mass basis by $\tilde
f_a$, as 
opposed to $\tilde f_i$, which we used to designate sfermions in 
the basis where the Yukawa
matrices (fermion mass matrices) are diagonal. 
\begin{equation}
  \label{eq:scalar-basis}
  \tilde{f}_a = U_{f_{a i}}^\dag \; \tilde f_i
\end{equation}

\subsection{Gaugino-sfermion-fermion Interactions}
\label{app:lagrangian}

There are two sources of fermion-sfermion-gaugino interactions. First,
these interactions arise in the superfield kinetic terms as a
consequence of  
gauge-invariance. Therefore these interactions have gauge-coupling
strength and couple gauginos to the fermions and the sfermions. The
sfermions and the fermions in these interactions have the same
chirality.  The Yukawa terms in the
superpotential are the second source of fermion-sfermion-gaugino
interactions. The superpotential interactions couple Higgsinos to
fermions and sfermions of opposite handedness.    

\subsubsection{Gluino-fermion-sfermion Interactions}

In the basis where all the fermions and as well as the sfermions are 
mass eigenstates the interactions are given as:
\begin{multline}
  - \sqrt 2 g_s \, \Bigg[ \bigg(  \bar{\lambda}_G \,
  \tilde u^*_{L_a} \left( U_{u_L} \right)_{\, a \, i} 
           \, P_L  -  \, \bar{ \lambda}_G^c  \,
  \tilde u^*_{R_a} \, \left( U_{u_R} \right)_{ a i} 
           \, P_R \bigg)  \, u_i   \\
   + \,  \bigg(  \bar{ \lambda}_G \,
  \tilde d^*_{L_a} \, \left( U_{d_L} \right)_{ a i} 
           \, P_L - \overline{ \lambda}_G^c \,
  \tilde d^*_{R_a} \, \left( U_{d_R} \right)_{ a i} 
          \, P_R \bigg) \, d_i \,  \Bigg]   \: + \: \text{c.c.} \; ,
\end{multline}
where we have abbreviated $\lambda_G  \equiv  \lambda_G^a  \, t_G^a$ 
and the matrices $t^a_G$ are the generators for the color group. 

\subsubsection{Neutralino-sfermion-fermion Interactions}

This set of interactions are slightly more complicated because 
we need to take into account not only the rotations that take the 
scalars to their mass eigenstates, but also to change basis 
for the neutralinos.  Resisting the temptation to write the final 
expression, we first write the interactions with neutralinos in 
the electroweak basis and sfermions in the fermion mass basis 
({\it i.e.} in terms of $f_i$).  Specifically, our expression below
is in terms of the vector $N$, which can be thought of a column of 
Dirac fermions.
We will also use four projection vectors $P_1, \dots , P_4$ that 
project out corresponding components. For example, 
$\bar \lambda_B =  \bar N P_1 $ - here $\bar N$ is a row as shown in 
Eq.~\eqref{mass-nut} and $P_1$ is a column with  zero elements
everywhere except at the first row, which is $1$. The
neutralino-fermion-sfermion interactions are now 
\begin{multline}
 - \sqrt 2 g \, \tan \theta_W \,  \Bigg[ 
              \overline N P_1 \left(  
      \frac{1}{6} \, \tilde u_{L_i}^* \,\delta_{i j} \, P_L \, u_j 
     +\frac{1}{6} \, \tilde d_{L_i}^* \,\delta_{i j} \, P_L \, d_j
     -\frac{1}{2} \, \tilde e_{L_i}^* \,\delta_{i j} \, P_L \, e_j
     -\frac{1}{2} \, \tilde \nu_{L_i}^* \,\delta_{i j} \, P_L \, \nu_j
              \right)   \\
             + \overline{ N^c} P_1 \left(
     -\frac{2}{3} \, \tilde u_{R_i}^* \,\delta_{i j} \, P_R \, u_j
     +\frac{1}{3} \, \tilde d_{R_i}^* \,\delta_{i j} \, P_R \, d_j
     +  \tilde e_{R_i}^* \,\delta_{i j} \, P_R \, e_j
            \right) \Bigg] \\ 
- \sqrt 2 g \, \Bigg[ \overline{ N} P_2 \, \Big(
      \frac{1}{2} \, \tilde u_{L_i}^* \,\delta_{i j} \, P_L \, u_j
     -\frac{1}{2} \, \tilde d_{L_i}^* \,\delta_{i j} \, P_L \, d_j
     -\frac{1}{2} \, \tilde e_{L_i}^* \,\delta_{i j} \, P_L \, e_j
     +\frac{1}{2} \, \tilde \nu_{L_i}^* \,\delta_{i j} \, P_L \, \nu_j   
     \Big)  \Bigg]   \\
- \sqrt 2  \, \frac{m_t}{v \sin \beta} \, \Bigg[ \Big(
     \overline{ N} P_3  \ \tilde u_{L_i}^* \,\delta_{i 3} \, P_R  + 
     \overline{ N^c} P_3 \ \tilde u_{R_i}^* \,\delta_{i 3} \, P_L   
    \Big) \; u_3 \Bigg] \\
- \sqrt 2  \, \frac{m_b}{v \cos \beta} \, \Bigg[ \Big( 
     \overline{ N} P_4  \ \tilde d_{L_i}^* \,\delta_{i 3} \, P_R  + 
     \overline{ N^c} P_4 \ \tilde d_{R_i}^* \,\delta_{i 3} \, P_L   
    \Big) \; d_3 \Bigg] \\
- \sqrt 2  \, \frac{m_\tau}{v \cos \beta} \, \Bigg[ \Big( 
     \overline{ N} P_4  \ \tilde e_{L_i}^* \,\delta_{i 3} \, P_R  + 
     \overline{ N^c} P_4 \ \tilde e_{R_i}^* \,\delta_{i 3} \, P_L   
    \Big) \; e_3 \Bigg] 
    \: + \:  \text{c.c.}\; .
\label{nut-eq-ew}
\end{multline}
The advantage of using matrix notation is now
evident. Neutralinos  and scalar sfermions in their mass basis are
related to the ones in Eq.~\eqref{nut-eq-ew} by 
$\overline N = \overline \chi \left(  L_N P_L + R_N P_R \right)$ and 
$\tilde f_i^* = \tilde f_a^* \: U_{f_{\,a\, i}} $\@. 
The final expressions are:
\begin{multline}
 - \sqrt 2 g \, \tan \theta_W \,  \Bigg[ \;
             \overline{ \chi} L_N P_1 \, \Big(  
      \; \frac{1}{6} \, \tilde u_{L_a}^* \,\left( U_{\tilde u_L} 
               \right)_{\, a \, j}  P_L \, u_j 
      +\, \frac{1}{6} \, \tilde d_{L_a}^* \,\left( U_{\tilde d_L} 
               \right)_{\, a \, j}  P_L \, d_j   \\
      -\, \frac{1}{2} \, \tilde e_{L_a}^* \,\left(U_{\tilde e_L} 
               \right)_{\, a \, j}  P_L \, e_j 
      -\, \frac{1}{2} \, \tilde \nu_{L_a}^* \,\left( U_{\tilde \nu_L} 
               \right)_{\, a \, j}   P_L \, \nu_j  \Big) \\  
             + \; \overline{ \chi^c} L_N P_1 \, \Big(
      -\, \frac{2}{3} \, \tilde u_{R_a}^* \,\left( U_{\tilde u_R} 
               \right)_{\, a \, j}   P_R \, u_j 
      +\, \frac{1}{3} \, \tilde d_{R_a}^* \,\left( U_{\tilde d_R} 
               \right)_{\, a \, j}   P_R \, d_j  \\
      +\,     \tilde e_{R_a}^* \,\left( U_{\tilde e_R} 
               \right)_{\, a \, j}   P_R \, e_j
            \Big) \Bigg]  \\
- \sqrt 2 g \,  \Bigg[ \overline{ \chi} L_N P_2 \Big( 
      \, \frac{1}{2} \, \tilde u_{L_a}^* \,\left( U_{\tilde u_L} 
               \right)_{\, a \, j}  P_L \, u_j
     -\, \frac{1}{2} \, \tilde d_{L_a}^* \,\left( U_{\tilde d_L} 
               \right)_{\, a \, j} \, P_L \, d_j \\
     -\, \frac{1}{2} \, \tilde e_{L_a}^* \,\left( U_{\tilde e_L} 
               \right)_{\, a \, j} \, P_L \, e_j 
     +\, \frac{1}{2} \, \tilde \nu_{L_a}^* \,\left( U_{\tilde \nu_L} 
               \right)_{\, a \, j} \, P_L \, \nu_j  \Big) 
     \Bigg] \; \\ 
- \sqrt 2  \, \frac{m_t}{v \sin \beta} \, \Bigg[ \Big( \,
     \overline{ \chi} R_N P_3  \  \tilde u_{L_a}^*  \left( U_{\tilde u_L} 
               \right)_{\, a \, 3}  P_R  + 
     \overline{ \chi^c} R_N P_3 \  \tilde u_{R_a}^*  \left( U_{\tilde u_R} 
               \right)_{\, a \, 3}  P_L  \Big) \; u_3 \Bigg]\\
- \sqrt 2  \, \frac{m_b}{v \cos \beta} \, \Bigg[ \Big( 
     \overline{ \chi} R_N P_4  \  \tilde d_{L_a}^* \left( U_{\tilde d_L} 
               \right)_{\, a \, 3}   P_R  + 
     \overline{ \chi^c} R_N P_4 \ \tilde d_{R_a}^* \left( U_{\tilde d_R} 
               \right)_{\, a \, 3}  P_L   \Big) \; d_3 \Bigg]\\
- \sqrt 2  \, \frac{m_\tau}{v \cos \beta} \, \Bigg[ \Big( 
     \overline{ \chi} R_N P_4  \ \tilde e_{L_a}^* \left( U_{\tilde e_L} 
               \right)_{\, a \, 3} P_R  + 
     \overline{ \chi^c} R_N P_4 \ \tilde e_{R_a}^*  \left( U_{\tilde e_R} 
               \right)_{\, a \, 3}  P_L  \Big) \; e_3 \Bigg]
               \: + \:  \text{c.c.}\; .
\label{nut-eq-mass}
\end{multline}

\subsubsection{Chargino-sfermion-fermion Interactions}

Deriving the chargino-fermion-sfermion expressions is entirely
analogous to the previous 
exercise where we found the interactions with the neutralinos. Here we
just give the final results where all the fields involved are 
mass eigenstates. The chargino-quark-squark interactions
already violate flavor, even if the quark and squark mass matrices are 
diagonalized simultaneously, due to the CKM matrix $V$. Additional
rotations of the squarks changes the amount of flavor violation.
\begin{multline}
- \,  g \,  \Bigg[ \overline{ \chi^{-}} L_C P_3  \Big(
      \, \tilde u_{L_a}^* \,\left( U_{\tilde u_L} V^\dag 
               \right)_{\, a \, j}  P_L \, d_j
    + \, \tilde \nu_{L_a}^* \,\left( U_{\tilde \nu_L} 
               \right)_{\, a \, j}  P_L \, e_j \Big) \\  
                  +\, \overline{ \chi^{+}} R_C P_1  \Big(
      \, \tilde d_{L_a}^* \,\left( U_{\tilde d_L} V
               \right)_{\, a \, j}  P_L \, u_j                  
      + \, \tilde e_{L_a}^* \,\left( U_{\tilde e_L} 
               \right)_{\, a \, j}  P_L \, \nu_j 
                     \Big) \Bigg]  \\
+ \sqrt 2  \, \frac{m_t}{v \sin \beta} \, \Bigg[ \,
     \overline{ \chi^{+}} L_C P_2  \  \tilde d_{L_a}^*  
     \left( U_{\tilde d_L} V \right)_{\, a \, 3}  P_R \,u_3   
   + \, \overline{ \chi^{-}} L_C P_3 \ \tilde u_{R_a}^* 
     \left( U_{\tilde u_R} V^\dag  \right)_{\, a \, 3}  P_L\, d_3 \Bigg] \\
+ \sqrt 2  \, \frac{m_b}{v \cos \beta} \, \Bigg[ \,
  \overline{ \chi^{-}} R_C P_4  \ \tilde u_{L_a}^* 
  \left( U_{\tilde u_L} V^\dag  \right)_{\, a \, 3}   P_R \, d_3 + \,
  \overline{ \chi^{+}} R_C P_4 \ \tilde d_{R_a}^* 
  \left( U_{\tilde d_R} V  \right)_{\, a \, 3}  P_L \, u_3   \Bigg] \\
+ \sqrt 2  \, \frac{m_\tau}{v \cos \beta} \, \Bigg[ 
  \overline{ \chi^{-}} R_C P_4  \ \tilde \nu_{L_a}^* 
  \left( U_{\tilde \nu_L}  \right)_{\, a \, 3} P_R \, e_3  + \,
  \overline{ \chi^{+}} R_C P_4 \ \tilde e_{R_a}^*  
  \left( U_{\tilde e_R} \right)_{\, a \, 3}  P_L\, \nu_3 \Bigg]  
  \: + \:  \text{c.c.}\; .
\label{charge-eq-mass}
\end{multline}

\end{appendix}


\begin{thebibliography}{99}

\bibitem{Ellis:1981ts}
  J.~R.~Ellis and D.~V.~Nanopoulos,
  Phys.\ Lett.\  B {\bf 110}, 44 (1982).

\bibitem{Donoghue:1983mx}
  J.~F.~Donoghue, H.~P.~Nilles and D.~Wyler,
  Phys.\ Lett.\  B {\bf 128}, 55 (1983).

\bibitem{Gabbiani:1996hi}
  For an older review, see 
  F.~Gabbiani, E.~Gabrielli, A.~Masiero and L.~Silvestrini,
  Nucl.\ Phys.\  B {\bf 477}, 321 (1996)
  [arXiv:hep-ph/9604387].

\bibitem{Kribs:2007ac}
  G.~D.~Kribs, E.~Poppitz and N.~Weiner,
  Phys.\ Rev.\  D {\bf 78}, 055010 (2008)
  [arXiv:0712.2039 [hep-ph]].

\bibitem{Blechman:2008gu}
  A.~E.~Blechman and S.~P.~Ng,
  JHEP {\bf 0806}, 043 (2008)
  [arXiv:0803.3811 [hep-ph]].

\bibitem{Plehn:2008ae}
  T.~Plehn and T.~M.~P.~Tait,
  arXiv:0810.3919 [hep-ph].

\bibitem{Bernreuther:2008ju}
  For a review, see e.g.\ W.~Bernreuther,
  J.\ Phys.\ G {\bf 35}, 083001 (2008)
  [arXiv:0805.1333 [hep-ph]].

\bibitem{Tait:2000sh}
  T.~M.~P.~Tait and C.~P.~P.~Yuan,
  Phys.\ Rev.\  D {\bf 63}, 014018 (2001)
  [arXiv:hep-ph/0007298].

\bibitem{Liu:2004bb}
  J.~J.~Liu, C.~S.~Li, L.~L.~Yang and L.~G.~Jin,
  Nucl.\ Phys.\  B {\bf 705}, 3 (2005)
  [arXiv:hep-ph/0404099].

\bibitem{Guasch:2006hf}
  J.~Guasch, W.~Hollik, S.~Penaranda and J.~Sola,
  Nucl.\ Phys.\ Proc.\ Suppl.\  {\bf 157}, 152 (2006)
  [arXiv:hep-ph/0601218].

\bibitem{Eilam:2006rb}
  G.~Eilam, M.~Frank and I.~Turan,
  Phys.\ Rev.\  D {\bf 74}, 035012 (2006)
  [arXiv:hep-ph/0601253].

\bibitem{Cao:2007dk}
  J.~J.~Cao, G.~Eilam, M.~Frank, K.~Hikasa, G.~L.~Liu, I.~Turan and J.~M.~Yang,
  Phys.\ Rev.\  D {\bf 75}, 075021 (2007)
  [arXiv:hep-ph/0702264].

\bibitem{Bozzi:2007me}
  G.~Bozzi, B.~Fuks, B.~Herrmann and M.~Klasen,
  Nucl.\ Phys.\  B {\bf 787}, 1 (2007)
  [arXiv:0704.1826 [hep-ph]].

\bibitem{LopezVal:2007rc}
  D.~Lopez-Val, J.~Guasch and J.~Sola,
  JHEP {\bf 0712}, 054 (2007)
  [arXiv:0710.0587 [hep-ph]].

\bibitem{Bejar:2008ub}
  S.~Bejar, J.~Guasch, D.~Lopez-Val and J.~Sola,
  Phys.\ Lett.\  B {\bf 668}, 364 (2008)
  [arXiv:0805.0973 [hep-ph]].

\bibitem{Datta:1997us}
  A.~Datta, J.~M.~Yang, B.~L.~Young and X.~Zhang,
  Phys.\ Rev.\  D {\bf 56}, 3107 (1997)
  [arXiv:hep-ph/9704257].

\bibitem{Oakes:1997zg}
  R.~J.~Oakes, K.~Whisnant, J.~M.~Yang, B.~L.~Young and X.~Zhang,
  Phys.\ Rev.\  D {\bf 57}, 534 (1998)
  [arXiv:hep-ph/9707477].

\bibitem{Berger:1999zt}
  E.~L.~Berger, B.~W.~Harris and Z.~Sullivan,
  Phys.\ Rev.\ Lett.\  {\bf 83}, 4472 (1999)
  [arXiv:hep-ph/9903549].

\bibitem{Berger:2000zk}
  E.~L.~Berger, B.~W.~Harris and Z.~Sullivan,
  Phys.\ Rev.\  D {\bf 63}, 115001 (2001)
  [arXiv:hep-ph/0012184].

\bibitem{Giudice:1998bp}
  For a review, see
  G.~F.~Giudice and R.~Rattazzi,
  Phys.\ Rept.\  {\bf 322}, 419 (1999)
  [arXiv:hep-ph/9801271].

\bibitem{Cohen:2006qc}
  A.~G.~Cohen, T.~S.~Roy and M.~Schmaltz,
  JHEP {\bf 0702}, 027 (2007)
  [arXiv:hep-ph/0612100].

\bibitem{Roy:2007nz}
  T.~S.~Roy and M.~Schmaltz,
  Phys.\ Rev.\  D {\bf 77}, 095008 (2008)
  [arXiv:0708.3593 [hep-ph]].

\bibitem{Murayama:2007ge}
  H.~Murayama, Y.~Nomura and D.~Poland,
  Phys.\ Rev.\  D {\bf 77}, 015005 (2008)
  [arXiv:0709.0775 [hep-ph]].

\bibitem{Ball:2007zza}
  G.~L.~Bayatian {\it et al.}  [CMS Collaboration],
  J.\ Phys.\ G {\bf 34}, 995 (2007).

\bibitem{Harnik:2008uu}
  R.~Harnik and G.~D.~Kribs,
  arXiv:0810.5557 [hep-ph].

\bibitem{Adriani:2008zr}
  O.~Adriani {\it et al.},
  arXiv:0810.4995 [astro-ph].

\bibitem{Mangano:2002ea}
  M.~L.~Mangano, M.~Moretti, F.~Piccinini, R.~Pittau and A.~D.~Polosa,
  JHEP {\bf 0307}, 001 (2003)
  [arXiv:hep-ph/0206293].

\bibitem{Maltoni:2002qb}
  F.~Maltoni and T.~Stelzer,
  JHEP {\bf 0302}, 027 (2003)
  [arXiv:hep-ph/0208156].

\bibitem{Alwall:2007st}
  J.~Alwall {\it et al.},
  JHEP {\bf 0709}, 028 (2007)
  [arXiv:0706.2334 [hep-ph]].

\bibitem{Sjostrand:2006za}
  T.~Sjostrand, S.~Mrenna and P.~Skands,
  JHEP {\bf 0605}, 026 (2006)
  [arXiv:hep-ph/0603175].

\bibitem{PGS}
J.~Conway,
http://www.physics.ucdavis.edu/~conway/research/software/pgs/pgs4-general.htm

\bibitem{Brooijmans:2008se}
 G.~H.~Brooijmans {\it et al.},
 arXiv:0802.3715 [hep-ph].

\bibitem{Catani:2001cc}
  S.~Catani, F.~Krauss, R.~Kuhn and B.~R.~Webber,
  JHEP {\bf 0111}, 063 (2001)
  [arXiv:hep-ph/0109231].

\bibitem{Cheng:2008mg}
  H.~C.~Cheng, D.~Engelhardt, J.~F.~Gunion, Z.~Han and B.~McElrath,
  Phys.\ Rev.\ Lett.\  {\bf 100}, 252001 (2008)
  [arXiv:0802.4290 [hep-ph]].

\bibitem{Cheng:2007xv}
  H.~C.~Cheng, J.~F.~Gunion, Z.~Han, G.~Marandella and B.~McElrath,
  JHEP {\bf 0712}, 076 (2007)
  [arXiv:0707.0030 [hep-ph]].

\bibitem{Burns:2008va}
  M.~Burns, K.~Kong, K.~T.~Matchev and M.~Park,
  arXiv:0810.5576 [hep-ph].

\bibitem{Athanasiou:2006ef}
  C.~Athanasiou, C.~G.~Lester, J.~M.~Smillie and B.~R.~Webber,
  JHEP {\bf 0608}, 055 (2006)
  [arXiv:hep-ph/0605286].

\bibitem{Wang:2006hk}
  L.~T.~Wang and I.~Yavin,
  JHEP {\bf 0704}, 032 (2007)
  [arXiv:hep-ph/0605296].

\bibitem{Bowen:2005xq}
  M.~T.~Bowen,
  Phys.\ Rev.\  D {\bf 73}, 097501 (2006)
  [arXiv:hep-ph/0503110].

\bibitem{:1999fr}
  ATLAS detector and physics performance. Technical design report. 
  Vol. 2, (1999).

\bibitem{Prieur:1019876}
     D.~ Prieur,
ATL-PHYS-PUB-2007-010. ATL-COM-PHYS-2007-013.

\bibitem{Aad:2009wy}
  G.~Aad {\it et al.}  [The ATLAS Collaboration],
  arXiv:0901.0512.

\bibitem{Ambrosanio:1996jn}
  S.~Ambrosanio, G.~L.~Kane, G.~D.~Kribs, S.~P.~Martin and S.~Mrenna,
  Phys.\ Rev.\  D {\bf 54}, 5395 (1996)
  [arXiv:hep-ph/9605398].

\bibitem{Kribs:2008hq}
  G.~D.~Kribs, A.~Martin and T.~S.~Roy,
  JHEP {\bf 0901}, 023 (2009)
  [arXiv:0807.4936 [hep-ph]].

\bibitem{Amsler:2008zzb}
 C.~Amsler {\it et al.}  [Particle Data Group],
 Phys.\ Lett.\  B {\bf 667}, 1 (2008).

\bibitem{Nisati:1997gb}
  A.~Nisati, S.~Petrarca and G.~Salvini,
  Mod.\ Phys.\ Lett.\  A {\bf 12}, 2213 (1997)
  [arXiv:hep-ph/9707376].

\bibitem{Allanach:2001sd}
  B.~C.~Allanach, C.~M.~Harris, M.~A.~Parker, P.~Richardson and B.~R.~Webber,
  JHEP {\bf 0108}, 051 (2001)
  [arXiv:hep-ph/0108097].

\bibitem{Ellis:1006573}
      J.~Ellis, A.~ Raklev, and O.~Oye, 
   ATL-PHYS-PUB-2007-016. ATL-COM-PHYS-2006-093.

\bibitem{Giagu:2008im}
  S.~Giagu  [ATLAS Collaboration and CMS Collaboration],
  arXiv:0810.1453 [hep-ex].

\end{thebibliography}
\end{document}